\documentclass[a4paper,11pt]{article}
\usepackage[margin=1in]{geometry}
\usepackage{epsfig}
\usepackage{cite}
\usepackage{graphicx}
\DeclareGraphicsExtensions{%
    .png,.PNG,%
    .pdf,.PDF,%
    .jpg,.mps,.jpeg,.jbig2,.jb2,.JPG,.JPEG,.JBIG2,.JB2}
\usepackage{epstopdf}
\usepackage{color}
\UseRawInputEncoding
\usepackage[english]{babel}
\usepackage{tabularx}
\usepackage{xparse}
\usepackage{braket}
\usepackage{amsmath}
\DeclareMathOperator{\sgn}{sgn}
\usepackage{mathrsfs}
\usepackage{hyperref}
\usepackage{grfext}
\usepackage{comment}
\usepackage{amssymb}
\usepackage{caption}
\usepackage{mathtools}
\usepackage{subcaption}
\usepackage{algorithm2e}
\usepackage{float}
\usepackage{subcaption}
\usepackage{multirow}
\restylefloat{table}
\usepackage{booktabs}
\usepackage[dvipsnames]{xcolor}
\usepackage{tikz}
\usepackage{pdflscape}
\usepackage{lscape}
\usepackage{tablefootnote}
\date{\today}
\def\unit{\leavevmode\hbox{\small1\kern-3.6pt\normalsize1}}

\def\gtwid{\mathrel{\raise.3ex\hbox{$>$\kern-.75em\lower1ex\hbox{$\sim$}}}}
\def\ltwid{\mathrel{\raise.3ex\hbox{$<$\kern-.75em\lower1ex\hbox{$\sim$}}}}
\def\gev{{\rm \, Ge\kern-0.125em V}}
\def\tev{{\rm \, Te\kern-0.125em V}}

\def    \be            {\begin{equation}}
\def    \ee            {\end{equation}}
\def    \bea           {\begin{eqnarray}}
\def    \eea           {\end{eqnarray}}

\def\a{\alpha}
\def\b{\beta}

\def\d{\delta}
\def\n{\nu}

\def\nn{\nonumber}
\def\d{\delta}
\def\D{\Delta}
\def\s{\sigma}
\def\r{\rho}
\def\t{\theta}

 \normalsize

 \normalsize

\newcommand{\bmat}{\left(\begin{array}}
\newcommand{\emat}{\end{array}\right)}

\begin{document}
\renewcommand{\thefootnote}{\fnsymbol{footnote}}
\vspace{.3cm}
\title{\Large\bf Textures of Neutrino Mass Matrix from $S_4$-flavor Symmetry} \author
{  N. Chamoun$^{1,2,3}$\thanks{nidal.chamoun@hiast.edu.sy, chamoun@uni-bonn.de} and E. I. Lashin$^{4,5}$\thanks{slashin@zewailcity.edu.eg, elashin@ictp.it}
 \\\hspace{-0.cm}
 \footnotesize$^1$ Physics Department, HIAST, P.O.Box 31983, Damascus, Syria. \\
 \footnotesize$^2$ Institute of Theoretical Physics,
CAS, P.O. Box 2735, Beijing 100190, China \\
\footnotesize$^3$ CASP,
Antioch Syrian University, Maaret Saidnaya, Damascus, Syria \\
\footnotesize$^4$  Department of Physics, Faculty of Science, Ain Shams University, Cairo 11566,  Egypt.  \\
\footnotesize$^5$ CFP, Zewail City of Science and
Technology, 
 \footnotesize  6 October City, Giza 12578, Egypt.
  \\\hspace{-0.cm}
 }
\date{\today}
\maketitle
\begin{abstract}
We study a texture of neutrino mass matrix characterized by two constraints consisting of one equality and another antiequality between two elements corresponding to two pairs of the matrix entries. Amidst such textures, we limit our study to three patterns which were realizable assuming an $S_4$-symmetry within type II-seesaw scenario. Three such cases were found and studied: I ($M_{\n 22}=-M_{\n 33}$ \& $M_{\n 11}=+M_{\n 23}$), II ($M_{\n 11}=-M_{\n 33}$ \& $M_{\n 22}=+M_{\n 13}$) and III ($M_{\n 11}=-M_{\n 22}$ \& $M_{\n 33}=+M_{\n 12}$). We specify the role of unphysical phases in the definition of the textures under study which were tested against experimental constraints, and were found to accommodate data with both hierarchies allowed. However, switching off the unphysical phases allows only for inverted hierarchy, except for the texture III which allows also, albeit for a very narrow parameter space region, for normal ordering. We stress that the different phenomenologies when including/excluding unphysical phases stem from the different definitions of the texture one has to adopt in order to make it insensitive to unphysical phases, rather than to any `absent' physical effects of unphysical phases. In addition, we made clear how a texture definition should be independent of the PMNS parametrization. We present a complete phenomenological analysis of these three textures and justify analytically the resulting correlations. We detail the effect of the unphysical phases in diluting/deforming several correlations, which otherwise would have been ``clear".  Finally, we give theoretical realizations within seesaw type II scenarios for such textures.
\end{abstract}
\maketitle
{\bf Keywords}: Neutrino Physics; Flavor Symmetry;
\\
{\bf PACS numbers}: 14.60.Pq; 11.30.Hv;
\vskip 0.3cm \hrule \vskip 0.5cm'
\section{Introduction}
Understanding the flavor structure of neutrinos is one of the notable problems in theoretical physics. Many attempts were carried out in order to build the structure of the neutrino mass matrix and its flavor mixing mechanism. Adding flavor symmetries to the standard model (SM) lead to neutrino textures of special form: tri-bimaximal mixing (TBM)\cite{tbm}, one zero element \cite{Rodejohan, 0texture}, two zero elements \cite{Frampton_2002, Xing_2002}, vanishing minor \cite{Lashin_2008}, two vanishing subtraces \cite{Alhendi_2008, ismael_2022},  two equalities \cite{2=_indian}, hybrid of zero element and minor \cite{hybrid}.

Although the examination of specific textures of Majorana neutrino mass matrix is
a traditional approach to the flavor structure in the lepton sector, its main motivation, however,
remains based on its simplicity and predictive power. The less constraints are put to define the texture, the easier it can accommodate the data, and so the more limited its predictive power is. Any phenomenological approach to flavor puzzles is useful and acceptable only
when it gives simple relations and/or interesting predictions for observables with
a small number of free parameters, which are suggestive of some nontrivial
symmetries or other underlying dynamics.

In \cite{gautam_2021}, a flavor symmetry based on the nonabelian group $A_4$ was used, within type I+II seesaw scenarios, to predict $\t_{23}$ near $45^o$, whereas the flavor group $S_3 \times Z_2$ was used in \cite{garcia_2021} within type II seesaw scenario. We, in \cite{ismael_npb} (\cite{ismael_2021}), studied textures with one equality (antiequality),  and found realizations using abelian groups within type I \& II seesaw scenarios.

In \cite{ismael_npb}, we studied the role played by the unphysical phases ($\phi^{\mbox{\tiny unphys}}$) in the definition of any texture, and although these unphysical phases may appear in some sectors beyond SM, however, after all symmetry breakings, they can be absorbed by the charged lepton fields and are thus non-physical in any setup involving just SM augmented by neutrino masses. On the other hand, unphysical phases represent details of the neutrino mass matrix $M_\n$ that are relevant for both processes of the mass matrix diagonalization and the renormaliztion group equations for physical parameters contained in $M_\n$\cite{Casas}. Also, the work of \cite{Manash} delved into the intricate roles the unphysical phases play underscoring their importance in mass matrix textures. The authors of \cite{Adhikary} stressed the importance of the unphysical phases for the diagonalization, and by eliminating them could find relations involving only the physical parameters. However, they did not discuss the role played by the unphysical phases in the way a texture of a particular mathematical form is defined, a point we aim to address in this work.

 Being non physical, one might be tempted to expect that scanning over the unphysical phases should not lead to a phenomenology of the other physical parameters different from that when equating the unphysical phases to zero. This expectation does not hold for all textures, and that motivated us to study the link between the unphysical phases and the texture definition, where we realized that forcing the unphysical phases to vanish corresponded to a different texture, albeit having the same mathematical form,  from when including them, and thus no reason for the phenomenology of the physical parameters to remain the same. Thus, switching on/off the unphysical phases changes the definition of the texture, given by a certain mathematical form, with different phenomenological results, bearing in mind that this does not mean that the unphysical phases have physical effects, but rather that in order to define a texture in a consistent way, such that it is insensitive to the ``non-physical" unphysical phases and also is independent of the parametrization one chooses for the PMNS matrix, one has to adopt different definitions when including or dropping the unphysical phases, whence the different phenomenologies one obtains. Put it differently, studying a texture with or without unphysical phases corresponds actually to studying two different textures, and so it is normal to get different phenomenologies.

 We present three definitions for any texture characterized by one mathematical constraint on the $M_\n$ elements. The first definition, called ``mathematical", is characterized by just the mathematical constraint and thus is independent of the parametrization one takes for the PMNS diagonalizing matrix. However, it is not invariant under rephasing, i.e. it can be met for one matrix $M_\n$ whereas it fails for another physically equivalent matrix $M'_\n$ differing only in the unphysical phases from $M_\n$. The second definition, called ``specific", is to restrict the mathematical constraint to the slice of vanishing unphysical phases, in that a mass matrix $M_\n$ meets the definition if its physically equivalent matrix with vanishing unphysical phases $M^{\mbox{\tiny phys}}_\n$ satisfies the mathematical constraint. Although this definition is by construction rephasing-invariant, however it depends on the parametrization one takes for the PMNS matrix since a vanishing unphysical phases slice in one parametrization may not correspond to a constant, let alone zero, unphysical phases slice in another parametrization. Moreover, any model leading to the desired form of the mathematical constraint form does not in itself constitute a model realizing the texture unless one can check that with the form obtained the unphysical phases are vanishing. The third definition, called ``generalized" stating that a matrix meets the texture condition if one of its physically equivalent matrices satisfies the mathematical constraint, presents the consistent way to define a texture which is both rephasing-invariant and parametrization-independent.

The aim of this manuscript is to study textures characterized by one equality and one antiequality, with and without restricting to the vanishing unphysical phases slice.  Underscoring the significance of the unphysical phases in the texture definition and consequently in its phenomenology has been raised in recent studies \cite{gautam_2021,2=_indian,yue}. Moreover, most of the models, presented in the literature, leading to a specific form of the texture do not guarantee that this given form corresponds to vanishing unphysical phases, whence the need to adopt a definition not specific to the vanishing slice of unphysical phases. Furthermore, as said above, the unphysical phases are dependent on the PMNS parametrization, and so a consistent ``physical'' texture definition should also be parametrization--independent as well as being insensitive to the unphysical phases. Making clear which texture definition one should specify so that to be reparametrization--invariant is one motivation behind our work.

 In general, with two conditions on the symmetric $3\times 3$ matrix, we have 4 real constraints, and the studied texture will have eight ($12-4$) free parameters. However, restricting to vanishing unphysical phases reduces the corresponding parameter space to 5-dim submanifold, which makes the predictive power of the texture acceptable in that accommodating all the experimental data with such a limited number of parameters is not a trivial task.  There are ($15 \times 14$) such textures, and instead of doing a thorough study, we opt to seek models based on $S_4$ within type II seesaw scenario, and limit the study to textures realizable by such models. Actually, $S_4$-flavor symmetry was used in \cite{ismael_2022} and found to lead naturally to some textures of the desired form with two antiequalities. Here, for one equality and one antiequality texture, we find three patterns which are realizable by seesaw scenarios based on $S_4$-models:
\bea
\label{3-textures}
I :\,\, M_{\n 22}=-M_{\n 33} &\& & M_{\n 11}=+M_{\n 23}, \nn \\ II :\,\, M_{\n 11}=-M_{\n 33} &\& & M_{\n 22}=+M_{\n 13}, \nn  \\ III :\,\, M_{\n 11}=-M_{\n 22} &\& & M_{\n 33}=+M_{\n 12}
\eea
We follow in this study, in order to justify analytically the phenomenological results, a strategy adopted in \cite{ismael_2022} based on the fact that the solar mass squared difference ($\d m^2$) is much smaller than the atmospheric mass squared difference ($\Delta m^2$) depicted in the parameter ($R_\nu=\frac{\d m^2}{\Delta m^2} \approx 10^{-2}$). So, instead of studying the complicated analytical formulae representing the different correlations between the observables, we study analytically the corresponding correlations resulting from putting ($R_\nu=0$), and we checked numerically that both sets of correlations give very similar results. Upon switching on the unphysical phases, we get enlarged acceptable parameter space regions, and not only some `clear' correlations disappear becoming diluted, but both types of hierarchies are viable now whereas zero unphysical phases favoured strongly the inverted type. Furthermore, upon adopting the toughest bound on cosmological constraints, the three textures are ruled out at vanishing unphysical phases, whereas only pattern I at normal ordering survives when including the unphysical phases, which shows clearly the phenomenological role these phases can play for a particular texture under study.

The plan of the paper is as follows. We present the notations in section 2, and the texture definition in section 3, followed by a digression in section 4 on the role played by the unphysical phases in the texture definition and how they are related to the parametrization taken. In section 5, we set precisely  the checking strategy we adopt, and proceed in section 6 to study phenomenologically the three textures motivated by $S_4$-realizability. First, we restrict to vanishing unphysical phases slice, then study the general case where these phases interfere as well in the texture definition. We present in detail an $S_4$-realization for the case I in section 7, and we end up with conclusion and summary in section 8. Two appendices include the essential group theory formulae for $S_4$.

\section{Notations}
We work in the `flavor' basis, where the charged lepton mass matrix is diagonal, and so the observed neutrino mixing matrix comes entirely from the neutrino sector:
\begin{equation}
V^{\dagger}M_{\nu}V^{*}= M_\n^{\mbox{\tiny diag.}} \equiv \mbox{diag}\left(m_1,m_2,m_3\right),
\end{equation}
with ($m_{i}, i=1,2,3$) real positive neutrino masses. The lepton mixing matrix $V$ contains three mixing angles, three CP-violating phases and three unphysical phases. It can be written as a Dirac mixing matrix $U_\d$ (consisting of three mixing angles and a Dirac phase) pre(post)-multiplied with a diagonal matrix $P_\phi$ ($P^{\mbox{\tiny Maj.}}$) consisting of three (two Majorana) phases. Thus, we have
\bea
V = P_\phi\;U_\d\;P^{\mbox{\tiny Maj.}} &:& U_\d \; = R_{23}\left(\t_{23}\right)\; R_{13}\left(\t_{13}\right)\; \mbox{diag}\left(1,e^{-i\d},1\right)\; R_{12}\left(\t_{12}\right), \label{defOfU} \nn
\\
P_\phi=\mbox{diag}\left(e^{i\phi_1},e^{i\phi_2},e^{i\phi_3}\right)&,&P^{\mbox{\tiny Maj.}} = \mbox{diag}\left(e^{i\rho},e^{i\sigma},1\right),
\\   U_{\mbox{\tiny PMNS}} = U_\d\;P^{\mbox{\tiny Maj.}}&=&  \!\!\! \left ( \begin{array}{ccc} c_{12}\, c_{13} e^{i\rho} & s_{12}\, c_{13} e^{i\sigma}& s_{13} \\ (- c_{12}\, s_{23}
\,s_{13} - s_{12}\, c_{23}\, e^{-i\delta}) e^{i\rho} & (- s_{12}\, s_{23}\, s_{13} + c_{12}\, c_{23}\, e^{-i\delta})e^{i\sigma}
& s_{23}\, c_{13}\, \\ (- c_{12}\, c_{23}\, s_{13} + s_{12}\, s_{23}\, e^{-i\delta})e^{i\rho} & (- s_{12}\, c_{23}\, s_{13}
- c_{12}\, s_{23}\, e^{-i\delta})e^{i\sigma} & c_{23}\, c_{13} \end{array}  \!\!\! \right ),\nn
  \label{defv}
\eea
where $R_{ij}(\theta_{ij})$ is the rotation matrix through the mixing angle $\theta_{ij}$ in the ($i,j$)-plane, ($\delta,\rho,\sigma$) are three CP-violating phases, and, adopting the parameterization where the third column of $U_{\mbox{\tiny PMNS}}$ is real, we denote ($c_{12}\equiv \cos\theta_{12}, s_{12}\equiv \sin\theta_{12}, t_{12}\equiv \tan\theta_{12}...)$.

The neutrino mass spectrum is divided into two classes: Normal hierarchy ($\textbf{NH}$) where $m_{1}<m_{2}<m_{3}$, and Inverted hierarchy ($\textbf{IH}$) where $m_{3}<m_{1}<m_{2}$. The solar and atmospheric neutrino mass-squared differences, and their ratio $R_{\nu}$, are defined as follows.
\begin{equation}
\delta m^{2}\equiv m_{2}^{2}-m_{1}^{2},~~\Delta m^{2}\equiv\Big| m_{3}^{2}-\frac{1}{2}(m_{1}^{2}+m_{2}^{2})\Big|,     \;\;
R_{\nu}\equiv\frac{\delta m^{2}}{\Delta m^{2}}.\label{Deltadiff}
\end{equation}
with data indicating ($R_{\nu}\approx 10^{-2}\ll1$). Two parameters which put bounds on the neutrino mass scales, by the  nuclear experiments on beta-decay kinematics and
neutrinoless double-beta decay, are the
effective electron-neutrino mass:
\begin{equation}
\langle
m\rangle_e \; = \; \sqrt{\sum_{i=1}^{3} \displaystyle \left (
|V_{e i}|^2 m^2_i \right )} \;\; ,
\end{equation}
and the effective Majorana mass term
$\langle m \rangle_{ee} $:
\begin{equation} \label{mee}
\langle m \rangle_{ee} \; = \; \left | m_1
V^2_{e1} + m_2 V^2_{e2} + m_3 V^2_{e3} \right | \; = \; \left | M_{\n 11} \right |.
\end{equation}
Our choice of parametrization has the advantage of not showing the Dirac phase $\d$ in the effective mass term of the double beta decay \cite{xing_2001,xing_2002}. However, one should note the slice of vanishing unphysical phases in this parametrization does not correspond to `constant', let alone vanishing,  unphysical phases in other parametrizations \cite{ismael_npb}.

The Jarlskog rephasing invariant quantity is given by \begin{equation}\label{jg}
J = s_{12}\,c_{12}\,s_{23}\, c_{23}\, s_{13}\,c_{13}^2 \sin{\delta}
\end{equation} and its value being non-vanishing is a necessary requirement for the presence of CP violation.

Cosmological observations put bounds on the  `sum'
parameter $\Sigma$:
\be
\Sigma = \sum_{i=1}^{3} m_i.
\ee

The allowed experimental ranges of the neutrino oscillation parameters at  3$\sigma$ level with the best fit values are listed in Table(\ref{TableLisi:as}) \cite{de_Salas_2021}.
\begin{table}[h]
\centering
\scalebox{0.8}{
\begin{tabular}{cccc}
\toprule
Parameter & Hierarchy & Best fit &  $3 \sigma$ \\
\toprule
$\delta m^{2}$ $(10^{-5} \text{eV}^{2})$ & NH, IH & 7.50 &  [6.94,8.14] \\
\midrule
 \multirow{2}{*}{$\Delta m^{2}$ $(10^{-3} \text{eV}^{2})$} & NH & 2.51 &  [2.43,2.59] \\
 \cmidrule{2-4}
           & IH & 2.48 &  [2.40,2.57]\\
\midrule
$\theta_{12}$ ($^{\circ}$) & NH, IH & 34.30 &  [31.40,37.40] \\
\midrule
\multirow{2}{*}{$\theta_{13}$ ($^{\circ}$)}  & NH & 8.53 & [8.13,8.92] \\
\cmidrule{2-4}
    & IH & 8.58 &  [8.17,8.96]\\
\midrule
\multirow{2}{*}{$\theta_{23}$ ($^{\circ}$)}  & NH & 49.26 &  [41.20,51.33] \\
\cmidrule{2-4}
      & IH & 49.46 &  [41.16,51.25]   \\
\midrule
\multirow{2}{*}{$\delta$ ($^{\circ}$)}  & NH & 194.00 &  [128.00,359.00] \\
\cmidrule{2-4}
 & IH & 284.00 & [200.00,353.00]   \\
\bottomrule
\end{tabular}}
\caption{\footnotesize The experimental bounds for the oscillation parameters at 3$\sigma$-level, taken from the global fit to neutrino oscillation data \cite{de_Salas_2021} (the numerical values of $\Delta m^2$ are different from those in the reference which uses the definition $\Delta m^2 = \Big| \ m_3^2 - m_1^2 \Big|$ instead of Eq. \ref{Deltadiff}). Normal and Inverted Hierarchies are
respectively denoted by NH and IH}.
\label{TableLisi:as}
\end{table}

For the non-oscillation parameters, we adopt the upper limits, which are obtained by KATRIN and Gerda experiment for $m_{e}$ and $m_{ee}$ \cite{Aker_2019,Agostini_2019} . However, we adopt for $\Sigma$ the results of Planck 2018 \cite{Planck} from temperature information with low energy by using the simulator SimLOW.
\begin{equation}\label{non-osc-cons}
\begin{aligned}
\Sigma~~~~~&<0.54~\textrm{ eV},\\
 m_{ee}~~&<0.228~\textrm{ eV},\\
 m_{e}~~~&<1.1~\textrm{ eV}.
\end{aligned}
\end{equation}

Furthermore, regarding the neutrino-mass-sum parameter  $\Sigma$, there are tighter bounds. The most severe one can be found in \cite{Valentino_2021}
\bea 
\Sigma&<& 0.09~\textrm{ eV} \label{cosTB},
\eea
 where it is derived by  using data from Supernovae Ia luminosity distances. Other tight bounds are the strict constraint of Planck 2018 combining baryon acoustic oscillation data in $\Lambda CDM$ cosmology ($\Sigma<0.12~\textrm{ eV}$) \cite{Planck} and the PDG live bound ($\Sigma<0.2~\textrm{ eV}$) \footnote{https://pdglive.lbl.gov/DataBlock.action?node=S066MNS} originating from fits assuming various cosmological considerations. Essentially, we are inclined to prefer a more relaxed constraint for $\Sigma$ since we would like to give more weight to colliders' data compared to cosmological considerations in testing our particle physics model, and also because the tight bounds make use of cosmological assumptions which are far from being anonymous \cite{Chacko_2020}.

Actually, the bound on $\Sigma$ (Eq. \ref{cosTB}) is very restrictive, and for comparison-purposes we considered, after having done the numerical analysis with the looser bound of (Eq. \ref{non-osc-cons}), also the tough bound of (Eq. \ref{cosTB}) whose admissible parameter space forms a subset of that of the loose bound, and checked, as we shall see without carrying out afresh the phenomenological analysis, that all textures under study are ruled out since the larger admissible space, and -a fortiori- any subset of it, is not consistent with the tight bound, except for one case (pattern I in {\bf NH} ordering with non-vanishing unphysical phases). For this latter case, we redid the analysis and presented the corresponding correlation plots. We repeat however that there is no universal agreement on the tight bound as it originates only from cosmological measurements assuming neutrinos are stable on cosmological timescales, whereas in models where neutrinos decay after becoming non-relativistic, a bound as large as $0.9$ eV is still allowed by the data\cite{Chacko_2020}.

For simplification/clarity purposes regarding the analytical expressions, we from now on denote the mixing angles as follows.
\begin{equation}
\theta_{12}\equiv\theta_{x},~~\theta_{23}\equiv\theta_{y},~~\theta_{13}\equiv\theta_{z}.\label{redefine}
\end{equation}
However, we shall keep the standard nomenclature in the tables and figures for consultation purposes.

\section{Textures with one equality and one antiequality}
Let us assume the (anti-)equality is between the entries ($M_{\nu~ab}$) $M_{\nu~\a\b}$ and ($M_{\nu~ij}$) $M_{\nu~nm}$ :
\begin{align}
M_{\nu~ab}+M_{\nu~ij}=0,\nonumber\\
M_{\nu~\a\b}-M_{\nu~nm}=0.\label{Traceconds}
\end{align}
We write Eq. (\ref{Traceconds}) in terms of the V matrix elements, where we define $U=P_\phi U_\d$, as
\begin{align}
\sum_{k=1}^{3}&(U_{ak}U_{bk}+U_{ik}U_{jk})\lambda_{k}=0,\nonumber\\
\sum_{k=1}^{3}&(U_{\a k}U_{\b k}-U_{nk}U_{mk})\lambda_{m}=0,\label{conds}
\end{align}
where
\begin{equation}
\lambda_1=m_1e^{2i\rho},~~\lambda_{2}=m_2e^{2i\sigma},~~\lambda_3=m_3.
\end{equation}

By writing Eq. (\ref{conds}) in a matrix form, we obtain
\begin{equation}
\left( \begin {array}{cc} A_1&A_2\\ \noalign{\medskip}B_1&B_2\end {array}
 \right)\left( \begin {array}{c} \frac{\lambda_1}{\lambda_3}\\ \noalign{\medskip}\frac{\lambda_2}{\lambda_3}\end {array} \right)=-\left( \begin {array}{c} A_3\\ \noalign{\medskip}B_3\end {array} \right),
\end{equation}
where
\begin{align}
A_k=&U_{ak}U_{bk}+U_{ik}U_{jk}\nonumber\\
B_k=&U_{\a k}U_{\b k}-U_{nk}U_{mk},~~~~~m=1,2,3.\label{Coff}
\end{align}
By solving Eq. (\ref{conds}), we obtain
\begin{align}
\frac{\lambda_1}{\lambda_3}=&\frac{A_3B_2-A_2B_3}{B_1A_2-A_1B_2},\nonumber\\
\frac{\lambda_2}{\lambda_3}=&\frac{A_1B_3-A_3B_1}{B_1A_2-A_1B_2}.
\end{align}
Therefore, we get the mass ratios and the Majorana phases in terms of the mixing angles, the Dirac phase and the unphysical phases
\begin{align}
m_{13} \equiv \frac{m_1}{m_3}=&\bigg|\frac{A_3B_2-A_2B_3}{B_1A_2-A_1B_2}\bigg|,\nonumber\\
m_{23} \equiv \frac{m_2}{m_3}=&\bigg|\frac{A_1B_3-A_3B_1}{B_1A_2-A_1B_2}\bigg|,\label{ratio}
\end{align}
and
\begin{align}
\rho=&\frac{1}{2}\mbox{arg}\bigg(\frac{A_3B_2-A_2B_3}{B_1A_2-A_1B_2}\bigg),\nonumber\\
\sigma=&\frac{1}{2}\mbox{arg}\bigg(\frac{A_1B_3-A_3B_1}{B_1A_2-A_1B_2}\bigg).
\end{align}
Note that the $A$'s and $B$'s are complex parameters depending on ($\d, \phi_k, k=1,2,3$) through exponentials, and so their complex conjugates can be obtained simply by ($\d \rightarrow 2\pi - \d, \phi_k \rightarrow 2\pi - \phi_k$) keeping the mass ratios invariant.

The neutrino masses are written as
\begin{equation}
m_3=\sqrt{\frac{\delta m^2}{m_{23}^2-m_{13}^2}},~~m_1=m_3\times m_{13},~~m_2=m_3\times m_{23}.\label{spectrum}
\end{equation}
As we see, we have eight input parameters corresponding $(\theta_{x},\theta_{y},\theta_{z},\delta,\delta m^2, \phi_1, \phi_2, \phi_3)$, which together with four real constraints in Eq. (\ref{conds}) allows us to determine the twelve degrees of freedom in $M_{\nu}$.

\section{Unphysical phases}

\subsection{Unphysical phases and Texture definition}
In any theory, which -after symmetry breakings- leads to the SM augmented with massive neutrinos, the unphysical phases, in the low energy regime, can be absorbed and set equal to zero. Any $P$ phase re-definition of the neutrino ``gauge" fields $\n^g$ would reflect itself as a rephasing of $M_\n$:
\bea
\label{phasing-neutrino-fields}
\n^g \rightarrow P \n^g &\Rightarrow& M^\n \rightarrow P^\dagger M^\n P^\dagger
\eea
The neutrino matrix $M_\n$ as a symmetric complex matrix belongs to an 12-dim real manifold $\cal M$. There are no physical effects of the unphysical phases $P_\phi = \mbox{diag}\left(e^{-i\phi_1}, e^{-i\phi_2}, e^{-i\phi_3}\right)$, and so any consistent texture definition should be insensitive to these unphysical phases. By redefining the neutrino fields through phasing, one can do without these unphysical phases but one has to redefine $M_\n$ in its turn, because of the rule of Eq. (\ref{phasing-neutrino-fields}) which one can call a ``rephasing''. More concretely, and for a fixed parametrization (say the PDG one or our adopted parametrization), we have the unique decomposition for any matrix $M_\n$:
\bea
M_\n = VM_\n^{\mbox{\tiny diag.}}V^T &\Rightarrow&
P_\phi^* M^\n P_\phi^* = M^{\mbox{\tiny phys}}_\n = U_{\mbox{\tiny PMNS}} M_\n^{\mbox{\tiny diag.}} U_{\mbox{\tiny PMNS}}^T.\eea
The 12-param $M_\nu$ and the 9-param $M^{\mbox{\tiny phys}}_\nu $ are two neutrino mass matrices which have the same physics, and we can define an equivalence relation:
\bea \label{equivalence} M_\n \sim M_\n' &\Leftrightarrow \exists  \mbox{\small      phase matrix } P_\phi :& M_\n'=P^*_\phi.M_\n.P^*_\phi \nn \\ P = \mbox{diag} (e^{-i\phi_1}, e^{-i\phi_2},e^{-i\phi_3}) &\rightarrow &
M'_{\n ij}=e^{i(\phi_i  + \phi_j)} M_{\n ij}\eea
This has consequences on defining a texture characterizing a set of neutrino mass matrices.
\begin{itemize}
\item
One may be inclined to  define a texture just by introducing a certain constraint given by ($g(M_\n)=0$), where $g$ is a vector function on $\cal M$:
\bea \label{def 1}
\mbox{`Mathematical' def. :}&M_\n \in \mbox{texture} \Leftrightarrow g(M_\n)=0&
\eea
However, this (`Mathematical' def.) is not valid in general as it may be sensitive to unphysical phases and may not be rephasing-invariant.  To clarify this last point,
we start by a simple texture example, which is a one zero texture defined, say, by ($M_{\n11}=0$). We find that this texture definition is insensitive to the unphysical phases as we have also ($M'_{\n11}=0$), and this zero-texture is indeed rephasing-invariant.
However, if we take a texture definition given by ($M_{\n11}=M_{\n22}$), we see that ($M'_{\n11}=e^{2i\phi_1}M_{\n11}, M'_{\n22}=e^{2i\phi_2}M_{\n22}$), and so we get ($M'_{\n11} = e^{2i(\phi_1-\phi_2)}M'_{\n22} \neq M'_{\n22}$). Thus,  the texture definition is met for $M_\n$ whereas it is not met for  $M'_\n$. This situation applies to the texture studied in this work.

\item
The correct way to define a texture is to define it on the equivalence classes ${\cal M}/\sim$, in that two equivalent matrices either both satisfy the texture definition or both fail it, such that the definition would be invariant under ``rephasing". Here, two common ways to meet this \cite{grimus,ismael_npb}:
\bea \mbox{ `Generalized' def. :} &M_\n \in \mbox{texture} \Leftrightarrow& \exists M_\n' \sim M_\n: g(M_\n')=0 \label{def 2}\\
\mbox{ `Specific' def. :} &M_\n \in \mbox{texture} \Leftrightarrow&  g(M_\n^{\mbox{\tiny phys.}})=0 \label{def 3}
\eea
\end{itemize}
We have the following set inclusions:
\bea
\mbox{`Mathematical' def. } \subseteq \mbox{`Generalized' def. } \supseteq \mbox{`Specific' def. }
\eea
Although the two definitions (`Mathematical' def. \& `Specific' def.) are not related and give in general different phenomenologies for the same constraint given by $g$, however, since the existence of solutions corresponding to (`Specific' def.) implies that of (`Mathematical' def.), then as long as one sticks to the  situation where the studied correlations do not involve unphysical phases,  then (`Specific' def.) represents a special case of (`Mathematical' def.), and these correlations between the physical parameters usually get ``diluted" when going from (`Specific' def.) to (`Mathematical' def.). Likewise, as long as the studied correlations involve just the physical parameters, then these correlations become identical for (`Mathematical' def.) and (`Generalized' def.). As to the correlations involving unphysical parameters, and although those of (`Mathematical' def.) are in general strictly included and not equal to those of (`Generalized' def.), which like the (`Specific' def.) case are a union of straight lines parallel to the axis representing the unphysical phase and covering all its allowable values, however they are devoid of any physical meaning.

From a practical point of view, after having picked up a neutrino gauge basis from the infinitely many equivalent bases, one fixes the unphysical phases to zero and scan over the remaining physical parameters when one adopts (`Specific' def.) searching for acceptable points meeting the texture definition and satisfying the experimental constraints. Scanning over all parameters including the unphysical phases would correspond to (`Mathematical' def.) which is dependent on the choice of the gauge-basis, and hence  is not valid physically. Although this latter scanning when adopted for (`Generalized' def.) may lead to redundancy, in that a point would be unnecessarily tested while an equivalent point differing only in unphysical phases has hitherto passed the tests meeting the texture definition and satisfying the experimental constraints, but it is still exhaustive vis-a-vis the physical parameters, and we get by scanning over all the physical and unphysical parameters the full correlations of (`Generalized' def.) provided they involve only the physical parameters, which is the case of interest.

We stress also that all past studies (e.g. \cite{0texture,liu,han,rodejohan}, which restricted the analysis to the vanishing unphysical phases slice, should be looked at as being carried out within (`Specific' def.), otherwise their analysis would have been susceptible to weaknesses having picked up a subset of the admissible parameter space.

We repeat that we shall adopt the physically meaningful definitions of (`Generalized' def.) and (`Specific' def.), corresponding to whether or not one scans over the unphysical phases,  which lead to different phenomenologies originating from different texture definitions and not from `physicality' of unphysical phases.

\subsection{Parametrization and unphysical phases}
We end this section by noting that the unphysical phases are determined once one fixes the PMNS parametrization. To clarify this statement, let us consider the PDG-parametrization where the mixing matrix is given by:
\bea
\label{pdg}
&\footnotesize{
 U_\d^{{\mbox{\tiny PDG}}}=\left ( \begin{array}{ccc} c_{12}\, c_{13}  & s_{12}\, c_{13} & s_{13} e^{-i\delta} \\ - s_{12}\, c_{23}- c_{12}\, s_{23}
\,s_{13} \, e^{i\delta}   &  c_{12}\, c_{23}\,- s_{12}\, s_{23}\, s_{13} e^{i\delta}
& s_{23}\, c_{13}\, \\ s_{12}s_{23}- c_{12}\, c_{23}\, s_{13} e^{i\delta}   &  - c_{12}\, s_{23}\, - s_{12}\, c_{23}\, s_{13} e^{i\delta}
 & c_{23}\, c_{13} \end{array} \right )
,}\nn\\  &\footnotesize{P^{\mbox{\tiny Maj.}}_{{\mbox{\tiny PDG}}} =
\mbox{diag}\left(e^{i\r}, e^{i\s},1\right)}, \footnotesize{P^{\mbox{\tiny PDG}}_\phi =
\mbox{diag}\left(e^{i\phi_1}, e^{i\phi_2},e^{i\phi_3}\right)}
\eea
where all the angles should be considered with the suffix ``PDG''.
By requiring the invariance of the neutrino mass matrix with respect to parametrization\footnote{Assuming $\d=\d^{\mbox{\tiny PDG}}$ and $\t=\t^{\mbox{\tiny PDG}}: \t\equiv(\t_x,\t_y\t_z)$, we could directly deduce the other transformation rules by noting that $U_\d = \textrm{diag}(e^{i\d},1,1) U_\d^{\mbox{\tiny PDG}} \textrm{diag}(e^{-i\d},e^{-i\d},1)$.}, we
have\footnote{If the familiar form $P^{\mbox{\tiny Maj.}}_{\mbox{\tiny PDG}}= \textrm{diag}(1,e^{i\varphi_2/2},e^{i(\varphi_3+2\d)/2})$ is used, then one gets  ($\r = \r^{\mbox{\tiny PDG}}+\d = - \frac{\varphi_3}{2}, \s = \s^{\mbox{\tiny PDG}}+\d= \frac{\varphi_2-\varphi_3}{2}$) \cite{0texture}.}:
\bea
\label{transformation_parametrization}
&\d=\d^{\mbox{\tiny PDG}}, \t=\t^{\mbox{\tiny PDG}}: \t\equiv(\t_x,\t_y,\t_z), \nn \\
&\phi_1 = \phi_1^{\mbox{\tiny PDG}} - \d, \phi_2 = \phi_2^{\mbox{\tiny PDG}}, \phi_3 = \phi_3^{\mbox{\tiny PDG}}, \nn \\
 &\r = \r^{\mbox{\tiny PDG}}+\d,
\s = \s^{\mbox{\tiny PDG}}+\d,
\eea
So, switching between the adopted parameterization in the paper (call it Adopted) and the PDG parametrization amounts  (noting that all the other parameters, namely, the Dirac phase $\d$, the mixing angles and the other two unphysical phases, are identical in the two parametrizations) to:
\vspace{-0.2cm}
\bea
\mbox{Adopted parametrization: } \left(\phi_1, \s, \r \right) &,&
\mbox{PDG parametrization: } \left(\phi'_1=\phi_1+\d, \s'=\s-\d,\r'=\r-\d\right) \label{parametrization}
\nn \\\eea
Thus, we see that a vanishing unphysical phases slice in one parametrization does not correspond to a constant, nor -a fortiori- a vanishing, unphysical phases slice in another parametrization. A second requirement for a consistent texture definition is to be paramterization-independent.

The `Mathematical' def. (Eq. \ref{def 1}) would remain the same upon reparametrization. However, the `Specific' def.  of Eq. (\ref{def 3}), using the vanishing unphysical phases, changes, unlike the (`Generalized' def.), upon reparametrization since this vanishing slice is different from parametrization to another, and this is another reason to prefer the (`Generalized' def.).

To recapitulate, we contrast in Tab. (\ref{properties}) the three different definitions, and emphasize the three requirements (namely of rephasing-invariance, parametrization-independence and realizability in model building) that a physically plausible definition of a texture should satisfy.

\begin{table}[h]
\centering
\scalebox{0.8}{
\hspace{-3cm}
\begin{tabular}{c||c|c|c|c|c|}
\toprule
$M_\n \in$ texture $\Leftrightarrow$ &$\phi^{\mbox{\tiny unphys.}}$-invariance &  Parametrization independence &  Physicality & $\phi^{\mbox{\tiny unphys}}$ correlations & realizability \\
\toprule \hline
 $g(M_\n)=0$ (`Mathematical' def.) & $\times$ & $\surd$ &  $\times$ & not trivial & $\surd$\\
 $g(M^{\mbox{\tiny{phys}}}_\n)=0$ (`Specific' def.) & $\surd$ & $\times$ &  $\times$ & trivial & $\times$\\
 $\exists M_\n' \sim M_\n: g(M_\n')=0$ (`Generalized' def.) & $\surd$ & $\surd$ &  $\surd$ & trivial & $\surd$ \\
\bottomrule
\end{tabular}}
\caption{\footnotesize Properties of the three different texture definitions. $\phi^{\mbox{\tiny unphys}}$-invariance means that the definition is defined for the equivalence class of matrices, where $M_\n' \sim M_\n$ means that both matrices have the same 9 physical observables and where $M_\n^{\mbox{\tiny{phys}}}$ is equivalent to $M_\n$ but with vanishing $\phi^{\mbox{\tiny unphys}}$. Because $\phi^{\mbox{\tiny unphys}}$'s are sensitive to the PMNS paramterization, then `Physicality' requires both $\phi^{\mbox{\tiny unphys}}$-invariance and PMNS parametrization-independence. By realizability we mean whether the model leading to a texture of the specified form can embody or not the definition.}
\label{properties}
\end{table}

The ``mathematical" definition (first line) of Table (\ref{properties}), as it is defined at the level of $M_\n$ elements, is parametrization independent, and once we find a model leading to the desired texture definition ($g(M_\n)=0$) then one can claim realizability, however it is not rephasing invariant in general.  In order to obtain the solutions according to this definition, one needs to scan over the full parameter space including the unphysical phases.

The ``specific'' definition (second line) of Table (\ref{properties}), suitable for many past texture studies, imposed a mathematical constraint of the form ($g(M_\n)=0$) but only in the vanishing unphysical phases slice. In practice, in order to get the ``solutions'', i.e. the points in the parameter space meeting the experimental constraints and satisfying the texture definition, in all the 12-dim $M_\n$ space, one limits the study of the texture to the slice of vanishing unphysical phases, and then any obtained mass-matrix-solution would generate 3-dim many more phenomenologically equivalent mass matrices, differing only in unphysical phases, by just rephasing this obtained solution. Restricting, plausibly, the phenomenological analysis to physical parameters correlation plots, though, means that one can take the plots solely from the solutions in that vanishing slice.  Anyway, there are two drawbacks for this way of defining the texture. First, the definition is not parametrization independent. As we saw, a vanishing slice in ``PDG" parametrization would not correspond to vanishing slice in our ``Adopted" parametrization, and vice versa, so the definition would start by fixing first a parametrization, then one imposes the texture definition in the slice of vanishing unphysical phases according to the chosen parametrization. However, the texture under study per se depends thus on the chosen PMNS parametrization. The second drawback is that none of the past studies adopting this definition, which introduced models leading to the desired form of the texture, checked that with the form they obtained, the unphysical phases were vanishing, and thus, in our opinion, the presented realization models were not complete, and this definition does not meet the ``realizability" criterion. These remarks apply for many textures studied in the past, and in particular to the texture studied in this work.

The third line ``generalized" definition is both rephasing-invariant and parametrization-independent and does not need any additional checking once you find a realization method to impose the mathematical constraint defining the texture. In practice, it corresponds to the mathematical definition supplemented by rephasing, and so, restricting to the physical correlation plots, one can obtain its solutions, like the mathematical definition, by scanning over the unphysical phases as well as the other parameters.

\section{$R_\n$-zeros as a  checking strategy for viability}
The mixing angles and the Dirac phase allow to determine the parameter $R_\n$:
\bea
\label{Rnu}
R_\n &=& \frac{m_{23}^2 - m_{13}^2}{\left| 1-\frac{1}{2} (m_{13}^2 + m_{23}^2 ) \right|}=f(\t, \d, \phi) \approx 10^{-2},
\eea
where  $\t\equiv(\t_x, \t_y, \t_z), \phi\equiv(\phi_1,\phi_2,\phi_3)$.

The strategy says that the real world corresponding to ($R_\n \approx 10^{-2} \neq 0$) is quite near the slice corresponding to ($R_\n  = 0$). More precisely, putting ($R_\n \approx 10^{-2}$) leads to ($\d=\d^{\mbox{\tiny real}}(\t_x,\t_y,\t_z, \phi)$), whereas putting ($R_\n =0$) leads to ($\d=\d^{\mbox{\tiny non-real}}(\t_x,\t_y,\t_z, \phi)$), and our strategy states that ($\d^{\mbox{\tiny real}} \approx \d^{\mbox{\tiny non-real}}$). In practice, and since the experimentally allowable range of $\t_z$ is quite tight, one might fix it to its best fit value ($\t_z \approx 8.5^o$) and consider $\d$ as a function of ($\t_x, \t_y, \phi$). These correlations would, in their turn, affect all other correlations involving quantities depending on ($\t,\d$).

Thus, we shall look for zeros of ($m_{23}^2-m_{13}^2$), which are the zeros of $R_\n$ as well as those of $\d m^2$, since these zeros will play the decisive role in determining the correlations plots.  Within the non-real world, analytical formulae are easier to deal with, which gives us an approximate justification of many correlation plots. We should distinguish between three levels of precision. The first one corresponds to ``full'' correlations with no approximations at all, and with taking all constraints into consideration. Then come the ``exact'' correlations assuming $\d m^2=0$, followed by ``approximate'' correlations equating to zero the first order of the expansion in terms of $\t_z$ of $(m_{23}^2-m_{13}^2)$ as a function of $(\t_x, \t_y, \t_z, \phi, \d)$.

We saw that under ($\d(\phi) \rightarrow 2\pi - \d(\phi)$), the A's and B's in Eq. (\ref{Coff}) go to their complex conjugates in such a way that mass ratios of Eq.(\ref{ratio}), as well as $R_\n$, remain invariant. Thus, any conclusion based on studying the expression ($m_{23}^2 - m_{13}^2$) should stay intact under this change, and so the admissibility of the parameter space point $(\t,\d,\phi)$ leads to that of the point $(\t,2\pi-\d,2\pi-\phi)$. However, the new experimental constraints on $\d$ do not quite respect this symmetry, so this remark just justifies the $\d$-evenness of `old' correlations plots restricted to vanishing $\phi$, say before the long baseline experiments gave constraints on $\d$ in 2017 \cite{Capozzi},  carried out with no constraints on $\d$.

Another remark states that if the zeros of ($m_{23}^2-m_{13}^2$) imply ($m_{13} = 1$), then the pattern in question can not accommodate data. This comes because one can not here get the good order of magnitude of $10^{-3}$ for $|m_3^2 - \frac{1}{2}(m_1^2+m_2^2)|$, since, up to order  $10^{-5} \approx (m_2^2 - m_1^2)$, we have $m_3=m_1$ giving $\D m^2= \frac{1}{2} (m_2^2-m_1^2) = O(10^{-5}) \neq 10^{-3}$. In general, one can plug any expression resulting from imposing zeros of ($m_{23}^2 - m_{13}^2$) into the expression of $m_{13}$ to deduce the hierarchy type.
For a detailed study of this checking strategy, one can refer to \cite{ismael_2022}.

\section{Numerical results}

In this section, we introduce the numerical and analytical results for the three $S_4$-inspired textures with one equality and another one antiequality. As said before, we shall start by studying the textures according to `Specific' def.  (Eq. \ref{def 3}) limiting the analysis to the vanishing unphysical phases slice, then we shall take the general `Generalized' def.  (Eq. \ref{def 2}) and compare between the two resulting phenomenologies. In order to relate between the mass matrix elements between the two cases we have:
\bea
\label{massmatrixelementsphase}
{M_\n}_{ab} (\phi \neq 0) &=& {M_\n}_{ab} (\phi =0) e^{i (\phi_a + \phi_b)}
\eea
We state in Table (\ref{AsBsExpressions}) the analytical expressions of the A's and B's in all textures, noting that the corresponding expressions for `Specific' def.  are obtained just by putting to zero the unphysical phases.
\begin{table}[h]
\begin{center}
\begin{tabular} {c||c}
\hline
\hline
Texture I & $M_{22}+M_{33}=0$ \& $M_{11}-M_{23}=0$ \\
\hline
$A_1$ & $e^{2i \phi_2}\left(c_x s_y s_z + s_x c_y e^{-i\d} \right)^2 + e^{2i \phi_3}\left(-c_x c_y s_z + s_x s_y e^{-i\d} \right)^2$ \\
$A_2$ & $e^{2i \phi_2} \left(-s_x s_y s_z + c_x c_y e^{-i\d} \right)^2 + e^{2i \phi_3}\left(s_x c_y s_z + c_x s_y e^{-i\d} \right)^2$ \\
$A_3$ & $e^{2i \phi_2} s_y^2 c_z^2+ e^{2i \phi_3} c_y^2 c_z^2 $ \\
$B_1$ & $e^{2i \phi_1} c_x^2 c_z^2 + e^{i (\phi_2+\phi_3)}\left(c_x s_y s_z + s_x c_y e^{-i\d} \right) \left(-c_x c_y s_z + s_x s_y e^{-i\d} \right)$ \\
$B_2$ & $e^{2i \phi_1}s_x^2 c_z^2 + e^{i (\phi_2+\phi_3)}\left(-s_x s_y s_z + c_x c_y e^{-i\d} \right) \left(s_x c_y s_z + c_x s_y e^{-i\d} \right)$ \\
$B_3$ & $e^{2i \phi_1}s_z^2 -e^{i (\phi_2+\phi_3)}s_y c_y c_z^2$ \\
\hline
\hline
Texture II & $M_{11}+M_{33}=0$ \& $M_{22}-M_{13}=0$ \\
\hline
$A_1$ & $e^{2i \phi_1}c_x^2 c_z^2 + e^{2i \phi_3}\left(-c_x c_y s_z + s_x s_y e^{-i\d} \right)^2$ \\
$A_2$ & $e^{2i \phi_1}s_x^2 c_z^2 + e^{2i \phi_3}\left(s_x c_y s_z + c_x s_y e^{-i\d} \right)^2$ \\
$A_3$ & $e^{2i \phi_1}s_z^2 + e^{2i \phi_3}c_y^2 c_z^2$ \\
$B_1$ & $e^{2i \phi_2}\left(c_x s_y s_z + s_x c_y e^{-i\d} \right)^2 - e^{i (\phi_1+\phi_3)}\left(-c_x c_y s_z + s_x s_y e^{-i\d} \right) c_x c_z $ \\
$B_2$ & $e^{2i \phi_2}\left(-s_x s_y s_z + c_x c_y e^{-i\d} \right)^2 + e^{i (\phi_1+\phi_3)}\left(s_x c_y s_z + c_x s_y e^{-i\d} \right) s_x c_z$ \\
$B_3$ & $e^{2i \phi_2} s_y^2 c_z^2 - e^{i (\phi_1+\phi_3)}c_y s_z c_z $ \\
\hline
\hline
Texture III & $M_{11}+M_{22}=0$ \& $M_{33}-M_{12}=0$ \\
\hline
$A_1$ & $e^{2i \phi_1}c_x^2 c_z^2 + e^{2i \phi_2}\left(c_x s_y s_z + s_x c_y e^{-i\d} \right)^2$ \\
$A_2$ & $e^{2i \phi_1}s_x^2 c_z^2 + e^{2i \phi_2}\left(-s_x s_y s_z + c_x c_y e^{-i\d} \right)^2 $ \\
$A_3$ & $e^{2i \phi_1}s_z^2 + e^{2i \phi_2}s_y^2 c_z^2$ \\
$B_1$ & $e^{2i \phi_3}\left(-c_x c_y s_z +s_x s_y e^{-i\d} \right)^2 +e^{i (\phi_1+\phi_2)} \left(c_x s_y s_z + s_x c_y e^{-i\d} \right) c_x c_z$ \\
$B_2$ & $e^{2i \phi_3}\left(s_x c_y s_z + c_x s_y e^{-i\d} \right)^2 - e^{i (\phi_1+\phi_2)}\left(-s_x s_y s_z + c_x c_y e^{-i\d} \right) s_x c_z$ \\
$B_3$ & $e^{2i \phi_3} c_y^2 c_z^2 - e^{i (\phi_1+\phi_2)}s_y s_z c_z$ \\
\hline
\hline
\end{tabular}
 \end{center}
 \caption{The analytical expressions for the coefficients A's and B's in all $S_4$-inspired textures.}
\label{AsBsExpressions}
\end{table}
\subsection{Numerical Results for Vanishing unphysical phases}
As mentioned before, the parameter space in this case is 5-dimensional consisting of 9 physical parameters made of three eigen masses ($m_1, m_2, m_3$), three mixing angles ($\t_{12}\equiv \t_x,\t_{23}\equiv\t_y,\t_{13}\equiv\t_z$) and three phase angles ($\rho, \sigma, \d$), reduced to 5 free parameters when taking the two complex constraints into account. Normally, we take these free parameters as ($\t_x, \t_y, \t_z, \d$) and one mass scale, say $\d m^2$. We scan over these free parameters and for each pattern we get 3-$\s$ level correlation plots, that we try to justify using the strategy of previous section.

We throw N points of order $(10^7-10^{10})$ in the 5-dimensional parameter space ($\theta_{x},\theta_{y},\theta_{z},\delta,\delta m^2$), and check first the  hierarchy type through Eqs. (\ref{ratio},\ref{spectrum}). Then, we test the experimental bounds of $\Delta m^2$ besides those of Eq. (\ref{non-osc-cons}) to see which regions are experimentally allowable. As the experimental bounds are not the same, except for $\theta_{x}$ and $\delta m^2$, in the two hierarchy cases, so we have to repeat the sampling for each hierarchy case.
We find that the three textures are viable with {\bf IH} ordering, but the texture III allows also, albeit in a very tight region, for {\bf NH} ordering. The predictions are given in Table \ref{Predictions-Vanishing}.
\begin{landscape}
\begin{table}[h]
\begin{center}
\begin{tabular} {c||c||c||c|c}
\hline
\hline
Observable & Pattern I &
Pattern II &
\multicolumn{2} {c} {Pattern III} \\
&$M_{\n 22}=-M_{\n 33} \&  M_{\n 11}=+M_{\n 23}$&$M_{\n 11}=-M_{\n 33} \&  M_{\n 22}=+M_{\n 13}$&\multicolumn{2} {c} {$M_{\n 11}=-M_{\n 22} \&  M_{\n 33}=+M_{\n 12}$} \\
\hline
\hline
Hierarchy & {\bf IH} & {\bf IH} &  {\bf IH} &  {\bf NH}  \\
\hline
$\theta_{12}^{\circ}\equiv \t_x^o$ & $31.4 - 37.39$ & $31.4-37.39$ & $31.4-37.4$ & $31.4-37.4$  \\
\hline
$\theta_{23}^{\circ} \equiv \t_y^o$ &$\{41.7\}\cup 42.39-44.63 \cup 45.28-47.66$ &$41.16-51.23$&$41.17-51.25$& $49.26-50.06$ \\ \hline
$\theta_{13}^{\circ}\equiv \t_z^o$ &$8.17-8.96$&$8.17-8.96$&$8.17-8.96$& $8.13-8.92$\\ \hline
$\delta^{\circ}$ & $200-261 \cup 276-352$ &$206.65-245.62$&$266.3-347.5$& $128-268.4 \cup 333-358.4$\\ \hline
$\rho^{\circ}$ & $5.2-17.98 \cup 162-178$&$95.95-107.77$&$65.52-88.02$& $65.72-99.65$\\ \hline
$\sigma^{\circ}$ &$\{13.5\} \cup 22.63-84.65 \cup 104-159 \cup 164.1-165 $ &$18.03-44.91$&$127.3-164.3$&$50.13-148.1$ \\ \hline
$m_1{\mbox{(eV)}}$ &$0.051-0.168$&$0.0673-0.0806$&$0.0655-0.1572$& $0.0216-0.0995$  \\ \hline
$m_2{\mbox{(eV)}}$ &$0.051-0.169$&$0.0678-0.0811$&$0.0661-0.1575$&$0.0233-0.0999$\\ \hline
$m_3{\mbox{(eV)}}$ &$0.051-0.161$&$0.0464-0.0632$&$0.0436-0.1494$& $0.0542-0.1117$ \\ \hline
$m_e{\mbox{(eV)}}$ &$0.05-0.168$&$0.0671-0.0804$&$0.0653-0.1571$& $0.0235-0.100$ \\ \hline
$m_{ee}{\mbox{(eV)}}$ &$0.013-0.161$ &$0.0314-0.0386$&$0.0299-0.0815$& $0.0203-0.0549$ \\
\hline
$\Sigma{\mbox{(eV)}}$ &$0.1162-0.4989$&$0.1816-0.2248$&$0.1742-0.4353$& $0.0993-0.3112$ \\
\hline
\hline
 \end{tabular}
 \end{center}
 \caption{The various predictions for the ranges of the neutrino physical parameters for three viable $S_4$-inspired textures at 3-$\sigma$ level, assuming vanishing unphysical phases.}
\label{Predictions-Vanishing}
 \end{table}
\end{landscape}

We see here that adopting the cosmological tough bound (Eq. \ref{cosTB}) will rule out all the patterns, since $\Sigma$ in all cases is always larger than $0.09$ eV.

From Table \ref{Predictions-Vanishing}, neither $m_1$ for normal hierarchy nor $m_3$ for inverted hierarchy does approach a vanishing value, so no singular such textures can accommodate the data.

We introduce 15 correlation plots for each texture, in any allowed hierarchy type, generated from the accepted points of the neutrino physical parameters at the 3-$\sigma$ level. The first and second rows represent the correlations between the mixing angles and the CP-violating phases. The third row introduces the correlations amongst the CP-violating phases, whereas the fourth one represents the correlations between the Dirac phase $\delta$ and each of $J$, $m_{ee}$ and $m_2$ parameters respectively. The last row shows the degree of mass hierarchy plus the ($m_{ee}, m_2$) correlation.

In order to interpret the numerical results and for later convenience, we introduce the two dimensionless quantities $\d_3 m^2$ and $\Delta_3 m^2$ related to the solar and atmospheric mass squared differences as,
\bea
\label{dimless_delta}
\d_3 m^2 &\equiv &  \d m^2/m_3^2 =m_{23}^2-m_{13}^2,\nn \\
\Delta_3 m^2 &\equiv &  \Delta m^2/m_3^2 ={m_{23}^2-m_{13}^2 \over \left| 1  - {1\over 2} \,\left( m_{13}^2 + m_{23}^2\right)\right|},
\eea
and consequently $ R_\n $, as defined in Eq.(\ref{Deltadiff}), can be expressed as,
\bea 
R_\n &=& {\d_3 m^2 \over \Delta_3 m^2 }.
\label{dimless_Rnu}
\eea 
These two dimensionless quantities, $\d_3 m^2$ and $\Delta_3 m^2$ as defined in Eq.(\ref{dimless_delta}), can be possibly written as an expansion in $s_z$ when the exact expressions turn out to be too complicated. The zeros of $\d_3 m^2$ can be analyzed analytically and numerically. By assuming these zeros, we can justify the hierarchy type and the  distinguishing features of the correlations plots as explained in the previous section.

Finally, we reconstruct $M_{\nu}$ for each viable ordering from the best fit representative point at the 3-$\sigma$ level in the 5-dimensional parameter space.

\subsubsection{Texture I ($M_{22}+M_{33}=0$ \& $M_{11}-M_{23}=0$) }
The A's and B's are read from Table (\ref{AsBsExpressions}) after having put to zero all the $\phi$'s.
We get
\bea
\label{I_Taylor_dm2}
m_{23}^2-m_{13}^2 &=& \frac{c_{2y}^2}{c_{2x}}-\frac{s_{2x} s_{2y} c_\d c_{2y} (1+s_{2y}) s_z}{c_{2x}^2}  +\mathcal{O}(s_z^2)
\eea
The exact zeros of $\d m^2$ are determined from, 
\bea
\label{I_exact_dm2}
m_{23}^2-m_{13}^2 &=&  \frac{\mbox{Num}(m_{23}^2-m_{13}^2)}{\mbox{Den}(m_{23}^2-m_{13}^2)}:\nn \\
\mbox{Num}(m_{23}^2-m_{13}^2) &=& -c_z^2\{-s_{2x} c_{2y} c_\d s_z \left[c_z^2 (s_{2y}+3)-2 \right] +
c_{2x} \left[c_z^2 \left(-2(s_{2y}-1)+c_{2y}^2\right) + 2 (s_{2y}-1) \right]\} \nn \\
&=&  c_z^2 c_{2y} \{s_{2x} c_\d s_z \left[ c_z^2 (s_{2y}+3)-2 \right] - c_{2x}
\left[ c_z^2  \left( \frac{2c_{2y}}{1+s_{2y}}+ c_{2y}\right)- \frac{2c_{2y}}{1+s_{2y}}\right]\},
\eea
where "Num" ("Den") stands for  Numerator (Denominator), a convention which will used from now on through the manuscript. 

As to the parameter $R_\n$ we get:
\bea
\label{I_r}
R_\n &=& \frac{2 c_{2y}^2 c_{2x}}{\left|1-6s_x^2 c_x^2 +s_{2y} s_{2x}^2 + s_y^2 c_y^2 (2 s_{2x}^2 - 4) -2 s_{2y} s_{2x}^2 c_\d^2\right|}   +\mathcal{O}(s_z)
\eea
We see here that making ($\t_y \rightarrow \frac{\pi}{4}$) makes $R_\n \ll 1$, and we expect the range of $\t_y$ to be in a neighbourhood around $\pi/4$. However, for ($\t_y \approx \frac{\pi}{4} \mbox{ and }  \d=\frac{3 \pi}{2}$), we get $R_\n$ too large, so the region $\t_y \approx \frac{\pi}{4}, \d = 3\pi/2$ is excluded.

\begin{figure}[hbtp]
\hspace*{-4cm}
\includegraphics[width=24cm, height=16cm]{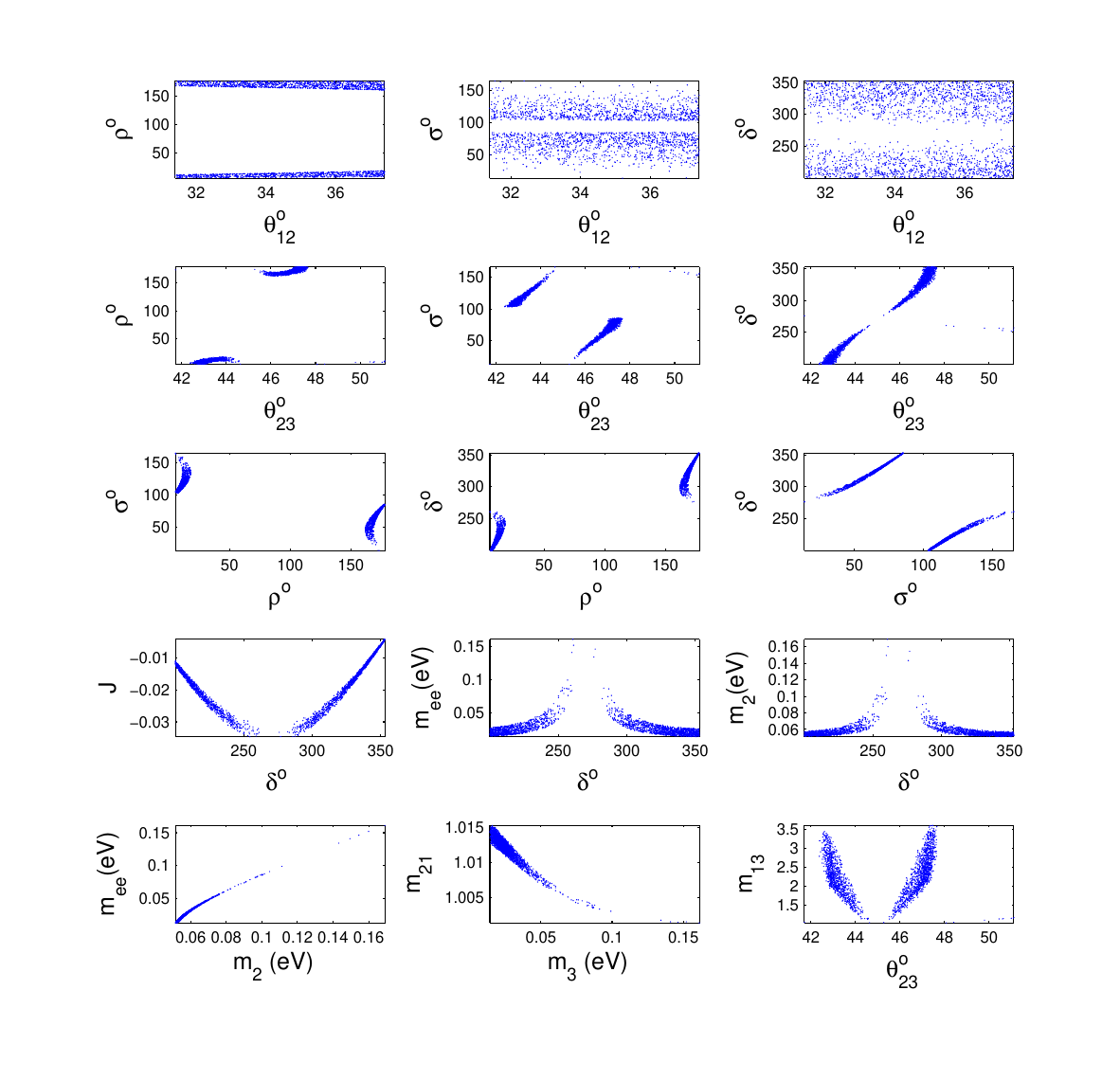}
\caption{The correlation plots for Texture I ($M_{22}+M_{33}=0$ \& $M_{11}-M_{23}=0$) at vanishing unphysical phases slice, in the inverted ordering hierarchy. The first and second row represent the correlations between the mixing angles ($\theta_{12}$,$\theta_{23}$) and the CP-violating phases. The third row introduces the correlations amdist the CP-violating phases, whereas the fourth one represents the correlations between the Dirac phase $\delta$ and each of $J$, $m_{ee}$ and $m_2$ parameters respectively. The last row shows the degree of mass hierarchy plus the ($m_{ee}, m_2$) correlation.}
\label{Iinv}
\end{figure}

Around ($\t_y = \frac{\pi}{4}, \t_z=0$), one can Taylor expand and get:
\bea
\label{Im2}
m_2^2 &=& \frac{\d m^2}{\t_y-\frac{\pi}{4}} \frac{c_{2x^2+c_\d^2s_{2x}^2}}{4 s_{2x}s_z c_\d}  +\mathcal{O}(1)
\eea
Thus the expression $\left[\left(\t_y -\frac{\pi}{4}\right) c_\d\right]$) should be positive, whence
\bea
\label{Im22}
\t_y <(>) \frac{\pi}{4} &\Rightarrow& \d <(>) \frac{3\pi}{2} \nn \\
\eea
This behaviour is observed numerically and is shown in the correlation plots of the Fig. (\ref{Iinv}), which show also a tendency of $m_2$ to blow up when $\d \rightarrow \frac{3\pi}{2}$, and so this latter value should be excluded, as we mentioned earlier.

As was stated before, studying the zeros of ($m_{23}^2-m_{13}^2$) would put us near the allowed points in the parameter space. From Eq. (\ref{I_exact_dm2}), we see that this corresponds to two regimes:
\begin{itemize}
\item $c_{2y} = 0 \Rightarrow \t_y = \frac{\pi}{4} $:

In this regime, expanding in powers of $(\t_y-\pi/4)$, we get the ``approximate" relation:
\bea
\label {I-ypi4-m13}
m_{13}&=& \sqrt{1+t_{2x}^2 c_\d^2} +\mathcal{O}(\t_y-\frac{\pi}{4})
\eea
This behaviour is confirmed by ``full'' numerical calculation of the correlation between $m_{13}$ and $\d$, and justified by the above formula and that the experimental values of $(\t_z, \t_y)$ are near the values ($0,45^o$). Moreover, we see that $m_{13}>1$ leading to an ordering of type {\bf IH} which can reach values up to $3.8$.

\item $s_{2x} c_\d s_z \left[ c_z^2 (s_{2y}+3)-2 \right] - c_{2x}
\left[ c_z^2  \left( \frac{2c_{2y}}{1+s_{2y}}+ c_{2y}\right)- \frac{2c_{2y}}{1+s_{2y}}\right] = 0 $:
 Substituting the zeros of this expression in the expression of $m_{13}^2$ we get for the allowed ranges:
 \bea
 \label {I-complic-m13}
 m_{13} &=& \sqrt{\frac{2(1-s_{2y})-c_{2y}^2 c_z^2}{c_{2y}^2 s_z^2}}
 \eea
One can plot this expression changing $\t_z$ and $\t_y$ in their respective allowed regions, and check one gets always {\bf IH}. 
\end{itemize}

Actually, we tested the $R_\n$-roots strategy by doing ``full'' scans over the free parameters, imposing by hand a constraint expressing the smallness of ($m_{23}^2 - m_{13}^2$ or of $R_\n$), and we checked that the resulting correlations are in a very good agreement with the ones of Fig. \ref{Iinv}

Finally, we reconstruct the neutrino mass matrix for a representative point, which is taken such that the mixing angles and the Dirac phase are chosen to be as near as possible from their best fit values from Table \ref{TableLisi:as}. For inverted ordering, the representative point is taken as following:
\begin{equation}
\begin{aligned}
(\theta_{12},\theta_{23},\theta_{13})=&(34.65^{\circ},45.79^{\circ},8.44^{\circ}),\\
(\delta,\rho,\sigma)=&(289.26^{\circ},167.22^{\circ},31.89^{\circ}),\\
(m_{1},m_{2},m_{3})=&(0.0754\textrm{ eV},0.0759\textrm{ eV},0.0571\textrm{ eV}),\\
(m_{ee},m_{e})=&(0.0569\textrm{ eV},0.0752\textrm{ eV}),
\end{aligned}
\end{equation}
the corresponding neutrino mass matrix (in eV) is
\begin{equation}
M_{\nu}=\left( \begin {array}{ccc} 0.0569 &  -0.0343 + 0.002i &  0.0353 - 0.002i\\ \noalign{\medskip}-0.0343 + 0.002i &   0.0089 &  0.0569
\\ \noalign{\medskip} 0.0353 - 0.002i &  0.0569 &  -0.0089\end {array} \right).
\end{equation}

\subsubsection{Texture II ($M_{11}+M_{33}=0$ \& $M_{22}-M_{13}=0$) }
\begin{figure}[hbtp]
\hspace*{-3.5cm}
\includegraphics[width=22cm, height=16cm]{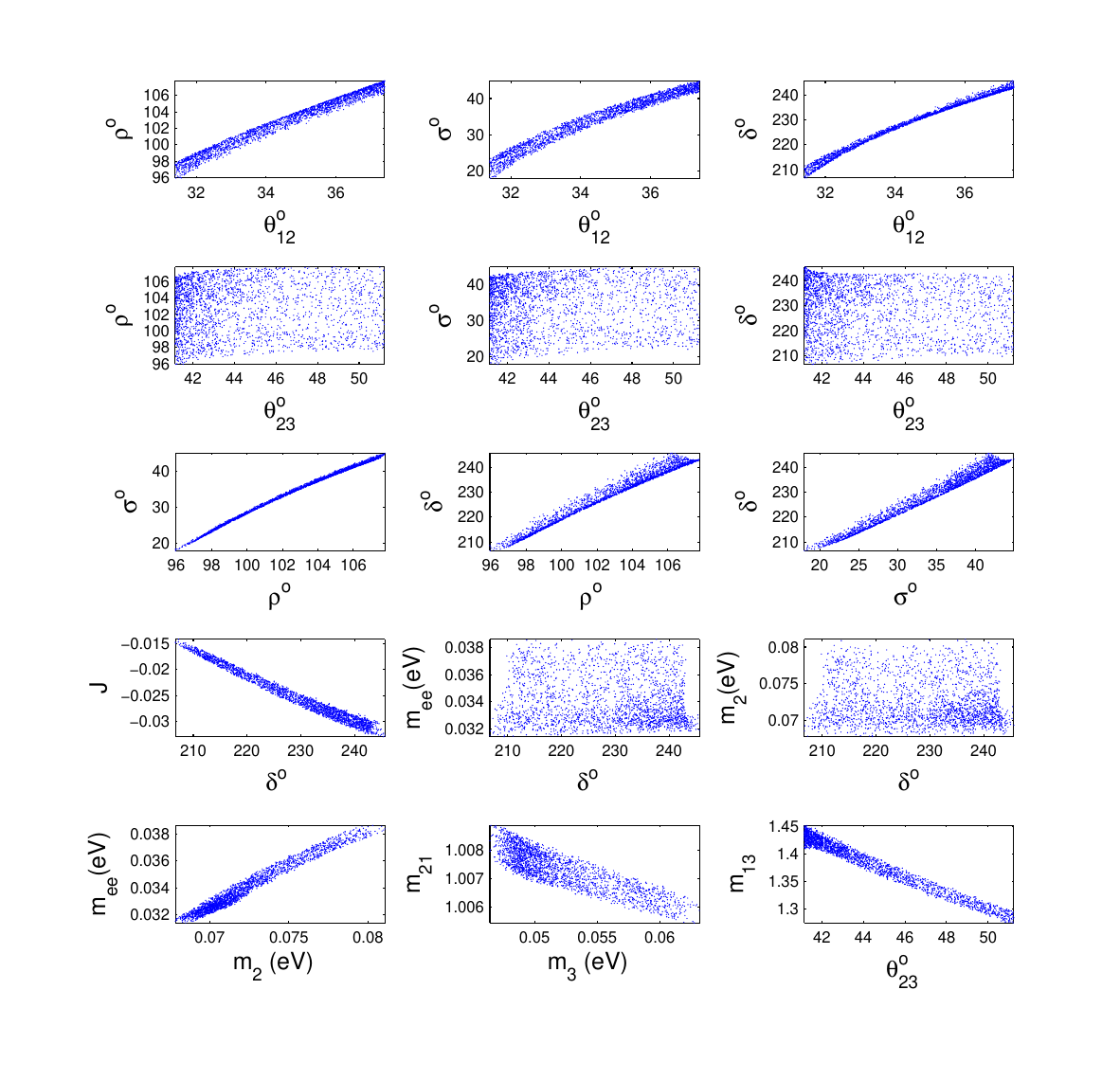}
\caption{The correlation plots for Texture II ($M_{11}+M_{33}=0$ \& $M_{22}-M_{13}=0$) at vanishing unphysical phases slice, in the case of inverted hierarchy. The first and second rows represent the correlations between the mixing angles $(\theta_{12},\theta_{23})$ and the CP-violating phases. The third and fourth rows show the correlations amidst the CP-violating phases and the correlations between the Dirac phase $\delta$ and each of $J$, $m_{ee}$ and $m_2$ parameters respectively. The last row shows the degree of mass hierarchy plus the ($m_{ee}, m_2$) correlation.}
\label{IIinv}
\end{figure}

The A's and B's are read from Table (\ref{AsBsExpressions}) after having put to zero all the $\phi$'s.
We get
\bea
\label{II_Taylor_dm2}
m_{23}^2-m_{13}^2 &=& \frac{Num}{Den}: \nn \\
Num &=& c_y^2 (3c_y^2 -2) (s_{2x} s_y c_\d + c_{2x}) +\mathcal{O}(s_z) \nn \\
Den &=& -4 s_x^2 c_x^2 s_y^4 c_\d^2 + s_{2x} s_y c_{2x} c_y^2 (c_y^2 -2) c_\d + \left( 1- (3+c_y^2) c_x^2 s_x^2 \right) c_y^4   +\mathcal{O}(s_z)
\eea
The exact zeros of $\d m^2$ are determined from
\bea
\label{II__exact_dm2}
m_{23}^2-m_{13}^2 &=&  \frac{\mbox{Num}(m_{23}^2-m_{13}^2)}{\mbox{Den}(m_{23}^2-m_{13}^2)}: \nn \\
\mbox{Num}(m_{23}^2-m_{13}^2) &=& s_y s_{2x} c_\d \left[-3 s_y^2 (1+c_y^2) c_z^5 + 2 s_z c_y (1+c_y^2) c_z^4 + (4-6c_y^4) c_z^3 - 2 s_z c_y (4-c_y^2) c_z^2 \right. \nn \\ && \left. + c_{2y} c_z + 2 s_z c_y \right]
 +c_{2x} \left[ (3-6c_y^2 -c_y^4) c_z^4 -2 s_z c_y s_y^2 (3c_y^2+1) c_z^3 \right. \nn \\ && \left. + (-2c_y^4-4+10c_y^2) c_z^2 + 2 s_z c_y s_y^2 c_z - c_{2y}\right] \label{II_Num_m232-m132}
\eea
As to the parameter $R_\n$, up to order $\mathcal{O}(s_z)$,  we get ($\sgn$ means the ``sign" function):
\bea
\label{II_r}
\mbox{Num}(\d_3 m^2) &=& c_y^2 (3c_y^2 -2) (c_{2x}+s_{2x}s_y c_\d) + \mathcal{O}(s_z) \nn \\
\mbox{Den}(\d_3 m^2) &=& -s_{2x}^2 s_y^4 c_\d^2 + \frac{1}{2} s_{2x} s_{2y} c_y c_{2x} (c_y^2 -2) c_\d -c_y^4 (-s_x^2 c_x^2 c_y^2 + 1 -3 c_x^2 s_x^2 ) +  \mathcal{O}(s_z) \nn \\
 \mbox{Num}(\D_3 m^2) &=& -8 c_x^2 s_x^2 c_y^2 s_y^2 c_\d^2 + s_{2x} s_y c_y^2 (3c_y^2 -4) c_{2x} c_\d + (3-10s_x^2c_x^2)c_y^4-6c_{2x}^2 c_y^2 + 2 c_{2x}^2 +  \mathcal{O}(s_z) \nn \\
 \mbox{Den}(\D_3 m^2) &=& 2\; \mbox{Den}(\d_3 m^2) \Rightarrow \nn \\
 R_\n &=& \frac{2 \; \mbox{Num}(\d_3 m^2) \; \sgn(\mbox{Den}(\d_3 m^2))}{\left|{\mbox{Num}}(\D_3 m^2) \right|} +  \mathcal{O}(s_z)
\eea
In order to check that one gets the correct ``full'' range of $\d$, i.e. $[206.65^o, 245.62^o]$ (look at Table \ref{Predictions-Vanishing}), assuming zeros of $R_\n$, we find that the ``exact" range, resulting from equating to zero the  expression of ($m^2_{23}-m^2_{13}$) (Eq. \ref{II_Num_m232-m132}), is $\d \in [205^o, 242.6^o]$ which is in good agreement with the ``full'' one. If want to compute the ``approximate" range of $\d$, originating from putting equal to zero the expansion of ($m^2_{23}-m^2_{13}$), we find that  one needs at least to go to first order expansion $(m^2_{23}-m^2_{13} \approx c_0 + c_1 s_z=0)$ so that there is a good agreement with the ``full" range. The correlation plots are shown in Fig. (\ref{IIinv}).

Again, imposing ($m_2=m_1$) (see Eq. (\ref{II_Num_m232-m132})), one finds the ``exact" expression:
\bea
m_{13}^2 &=& \frac{(1-c_z^2s_y^2)\left(1-(1+c_y^2)^2 c_z^4 + 4 c_y^2 c_z^2 -2 s_{2z}c_y\right)}{s_y^2 \left(1-(1+c_y^2)c_z^2\right)^2 c_z^2}  \label{II_exact_m13}
\eea
from whence we get an ``exact" estimation of ($m_{13} \in [1.264, 1.45]$), to be compared with the ``full" estimation ($m_{13} \in [1.273, 1.452]$), which shows a good agreement. However, had we expanded $m_{13}$ of Eq. (\ref{II_exact_m13}) in powers of $s_z$, then one needs to go to $O(s_z^2)$-term in order to get an ``approximate" good estimation ($m_{13} \in [1.307, 1.472]$). In all cases, the hierarchy type is found to be of inverted ordering.

Finally, we reconstruct the neutrino mass matrix for a representative point, chosen such that $\t,\d$ as near as possible from their best fit values. For inverted ordering, the representative point is taken as following:
\begin{equation}
\begin{aligned}
(\theta_{12},\theta_{23},\theta_{13})=&(34.53^{\circ},46.47^{\circ},8.59^{\circ}),\\
(\delta,\rho,\sigma)=&(229.79^{\circ},103.26^{\circ},35.92^{\circ}),\\
(m_{1},m_{2},m_{3})=&(0.0727\textrm{ eV},0.0732\textrm{ eV},0.0540\textrm{ eV}),\\
(m_{ee},m_{e})=&(0.0348\textrm{ eV},0.0725\textrm{ eV}),
\end{aligned}
\end{equation}
Note the destructive interference between the contributions of $m_1$ and $m_2$ in determining $m_{ee}$ (Look at at Eq. \ref{mee}). The corresponding neutrino mass matrix (in eV) is
\begin{equation}
M_{\nu}=\left( \begin {array}{ccc} -0.0348+0.0003i &  -0.0331 + 0.0004i &  0.0543 - 0.0004i\\ \noalign{\medskip}-0.0331 + 0.0004i &   0.0543-0.0004i &  0.0069+0.0004i
\\ \noalign{\medskip} 0.0543 - 0.0004i &  0.0069+0.0004i &  0.0348-0.0003i\end {array} \right).
\end{equation}

\subsubsection{Texture III ($M_{11}+M_{22}=0$ \& $M_{12}-M_{33}=0$) }
The A's and B's are read from Table (\ref{AsBsExpressions}) after having put to zero all the $\phi$'s.
We get
\bea
\label{III_Taylor_dm2}
m_{23}^2-m_{13}^2 &=& \frac{\mbox{Num}}{\mbox{Den}}: \nn \\
\mbox{Num} &=& s_y^2(3c_y^2-1) (s_{2x} c_y c_\d - c_{2x}) +\mathcal{O}(s_z) \nn \\
\mbox{Den} &=& -s_{2x}^2c_y^4c_\d^2+s_{2x}c_y s_y^2 (1+c_y^2) c_{2x}c_\d -s_y^4 (s_x^2 c_x^2c_y^2+c^2_{2x})   +\mathcal{O}(s_z)
\eea
The exact zeros of $\d m^2$ are determined from
\bea
\label{III__exact_dm2}
m_{23}^2-m_{13}^2 &=&  \frac{\mbox{Num}(m_{23}^2-m_{13}^2)}{\mbox{Den}(m_{23}^2-m_{13}^2)}: \nn \\
\mbox{Num}(m_{23}^2-m_{13}^2) &=&  s_{2x} c_y \left[ -3c_y^2(1+s_y^2)c_z^5+2s_ys_z(1+s_y^2)c_z^4 + (-2-6c_y^4+12c_y^2) c_z^3\right.\nn \\ && \left. -2s_ys_z(c_y^2+3)c_z^2-c_{2y}c_z+2s_ys_z\right]c_\d \nn\\
&& -c_{2x} \left[ (-c_y^4-4+8c_y^2) c_z^4 + 2c_y^2 s_y (3c_y^2-4)s_zc_z^3 \right. \nn \\ &&
\left. +(4-2c_y^4-6c_y^2)c_z^2+c_y^2s_ys_{2z}+1+2c_y^2\right] \label{III_Num_m232-m132}
\eea
As to the parameter $R_\n$, up to order $\mathcal{O}(s_z)$,  we get:
\bea
\label{III_r}
\mbox{Num}(\d_3 m^2) &=& -s_y^2(3c_y^2-1)(s_{2x}c_yc_\d-c_{2x}) + \mathcal{O}(s_z) \nn \\
\mbox{Den}(\d_3 m^2) &=& s_{2x}^2 c_y^4c_\d^2-s_{2x}c_ys_y^2(1+c_y^2)c_{2x}c_\d + s_y^4 (c_x^2 s_x^2 c_y^2 + c_{2x}^2) +  \mathcal{O}(s_z) \nn \\
 \mbox{Num}(\D_3 m^2) &=& \frac{1}{2}s_{2x}^2 s_{2y}^2c_\d^2-s_{2x}c_ys_y^2(3c_y^2+1)c_{2x}c_\d
 +(10c_x^2s_x^2-3)c_y^4+s^2_{2x}c_y^2-6s_x^2c_x^2+1
  +  \mathcal{O}(s_z) \nn \\
 \mbox{Den}(\D_3 m^2) &=& 2 \; \mbox{Den}(\d_3 m^2) \Rightarrow \nn \\
 R_\n &=& \frac{2\; \mbox{Num}(\d_3 m^2) \; \sgn(\mbox{Den}(\d_3 m^2)}{\left|\mbox{Num}(\D m^2) \right|} +  \mathcal{O}(s_z)
\eea
Assuming zeros for $\mbox{Num}(m_{23}^2-m_{13}^2)$ (Eq. \ref{III_Num_m232-m132}), one can get the "exact" mass ratio as:
\bea
\label{III_m13^2}
m_{13}^2 &=& \frac{(1-c_y^2 c_z^2)\left(1-(1+s_y^2)^2c_z^4+4s_y^2c_z^2-2s_ys_{2z}\right)}{(c_y^2c_z^2-c_{2z})^2c_y^2c_z^2}
\eea
leading to
\bea
\label{III_m13}
m_{13} &=& \sqrt{2+t_y^2} \left(1-\frac{2s_z}{s_y(1+c_y^2)}\right) +{\cal O}(s_z^2)
\eea
It is easy to plot, for a fixed $\t_z$, the graph $m_{13}(\t_y)$, using the leading up to $s_z$ term of Eq. (\ref{III_m13}), and check it is always above $m_{13}=1$. By changing $\t_z$ within its admissible interval $[8.13^o, 8.96^o]$, one can get thus an ``approximate" estimation of $m_{13} \in [1.16,1.4]$. On the other hand, the ``exact''  range of $m_{13}$ originating from equating $\mbox{Num}(m_{23}^2-m_{13}^2)$ (Eq. \ref{III__exact_dm2}) to zero is found to be $[1.297,1.502]$. Thus we deduce that the ordering is of type {\bf IH}.

Actually, by ``full'' numerics scanning, we found that some points, concentrated around $\t_y \approx 50^o$, present an ordering of type {\bf NH}. In order to interpret this fact analytically using the $R_\n$-zeros strategy, we note that  $\mbox{Num}(m_{23}^2-m_{13}^2)$ (Eq. \ref{III__exact_dm2}) has the following structure:
\bea
\mbox{Num}(m_{23}^2-m_{13}^2) &=& {\cal A}(\t_x,\t_y,\t_z) c_\d + {\cal B}(\t_x,\t_y,\t_z)
\eea
giving when imposed to be zero the analytical expression of ($c_\d=\frac{-{\cal B}}{{\cal A}}$) used in subsequent evaluation of ``exact" $m_{13}$. However, this process excludes the points corresponding to ${\cal A}={\cal B}=0$. Actually, we found that for each admissible $\t_z$, there is a quasi single value of $\t_y \approx 50^o$ (e.g. we get $\t_y=49.47^o (49.96^o)$ for $\t_z=8.13^o (8.96^o)$) where both ${\cal A}$ ad ${\cal B}$ are almost zero through the admissible range of $\t_x$. We thus interpret the ``bump" of {\bf NH} ordering around $\t_y=50^o$.

\begin{figure}[hbtp]
\includegraphics[width=11cm, height=8cm]{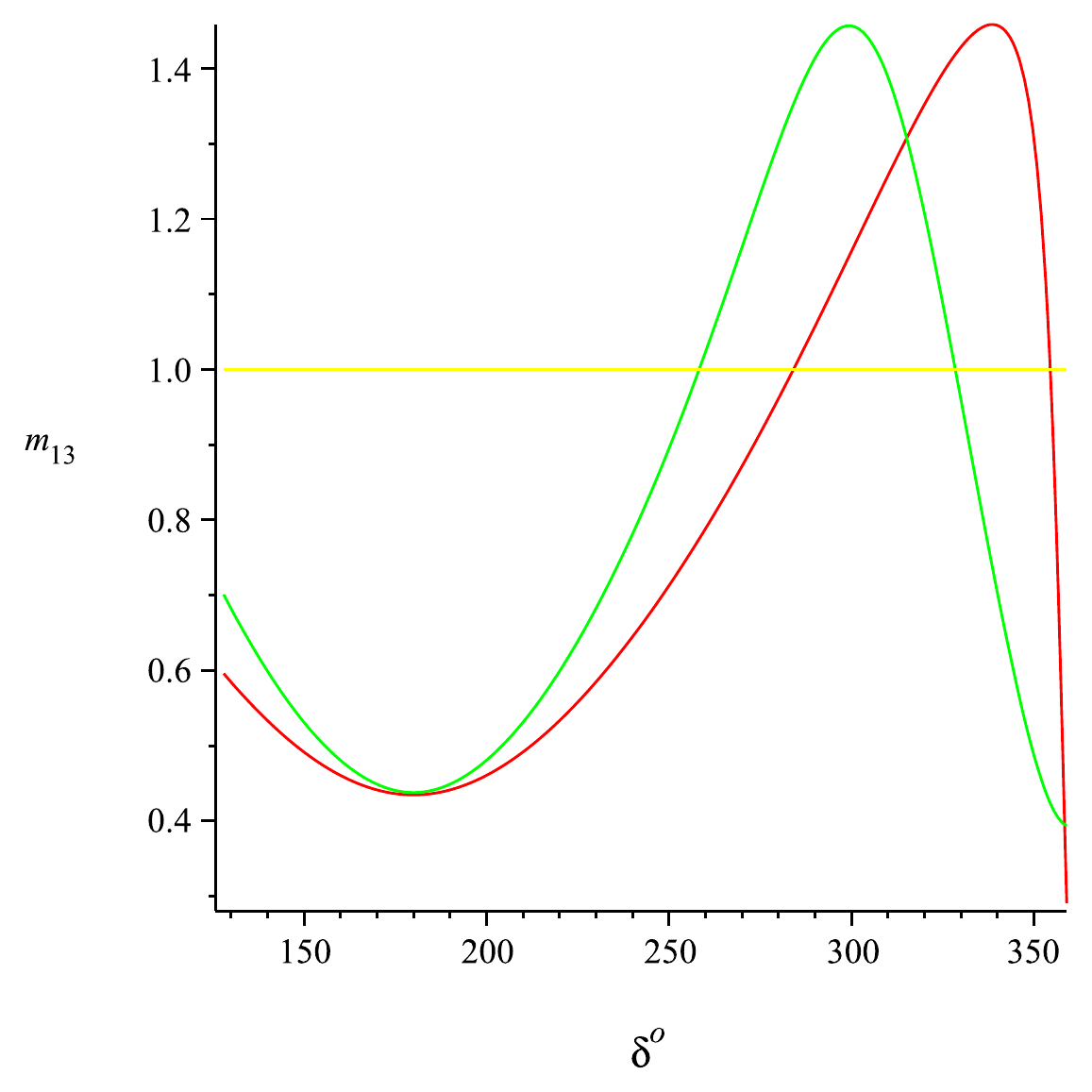}
\caption{Two types of hierarchy around $\t_y =50^o$. The graph $m_{13}(\d)$ for fixed ($\t_z=8.5^o$), representing the best fit, and ($\t_y=50^o$). For $\t_x$, we take its extremal admissible values $\t_x=31.4^o$ (red) or $37.4^o$ (green). For most regions of $\d$ the ordering is of type {\bf NH}, with a forbidden band ($[284^o, 328^o]$) for {\bf NH}, where both colored curves are higher than $1$.}
\label{m13d}
\end{figure}

Actually, fixing the narrowly changing $\t_z$ at its best fit ($8.5^o$), and setting $\t_y$ to $50^o$, we draw in (Fig. \ref{m13d}) the ``full'' graphs of ($m_{13}$) versus $(\d)$ taking for $\t_x$ its extremal admissible values of $31.4^o$ (red) and $37.4^o$ (green). We see that indeed two types of hierarchies ({\bf NH},{\bf IH}) are allowable, according to whether $m_{13}$ is smaller or larger than $1$, but whereas the {\bf IH} ordering requires ($\d > 258^o$), outside which both colored curves are lower than $1$, the {\bf NH} ordering has a forbidden gap for $[284^o, 328^o]$ for $\d$ where both colored curves are higher than $1$.

\begin{enumerate}

\item {\bf Texture III: Inverted Hierarchy}

As said above, this ordering occurs for $\t_y$ distributed through its admissible interval. The correlation plots are shown in Fig. (\ref{IIIinv}).

We reconstruct the neutrino mass matrix for a representative point which, for inverted ordering, is taken as follows (the parameters $\t, \d$ are chosen to be near their best fit values):
\begin{equation}
\begin{aligned}
(\theta_{12},\theta_{23},\theta_{13})=&(36.41^{\circ},49.49^{\circ},8.54^{\circ}),\\
(\delta,\rho,\sigma)=&(289.74^{\circ},77.81^{\circ},142.90^{\circ}),\\
(m_{1},m_{2},m_{3})=&(0.0726\textrm{ eV},0.0731\textrm{ eV},0.0534\textrm{ eV}),\\
(m_{ee},m_{e})=&(0.0343\textrm{ eV},0.0724\textrm{ eV}).
\end{aligned}
\end{equation}
Note the destructive interference between the contributions of $m_1$ and $m_2$ in determining $m_{ee}$ (Look at Eq. \ref{mee}). The corresponding neutrino mass matrix (in eV) is
\begin{equation}
M_{\nu}=\left( \begin {array}{ccc} -0.0339-0.0053i &  0.0479 + 0.0151i &  -0.0359 - 0.0164i\\ \noalign{\medskip} 0.0479 + 0.0151i  &   0.0339+0.0053i &  0.0118-0.0096i
\\ \noalign{\medskip} -0.0359 - 0.0164i &  0.0118-0.0096i &  0.0479+0.0151i\end {array} \right).
\end{equation}

\begin{figure}[hbtp]
\hspace*{-3.5cm}
\includegraphics[width=22cm, height=16cm]{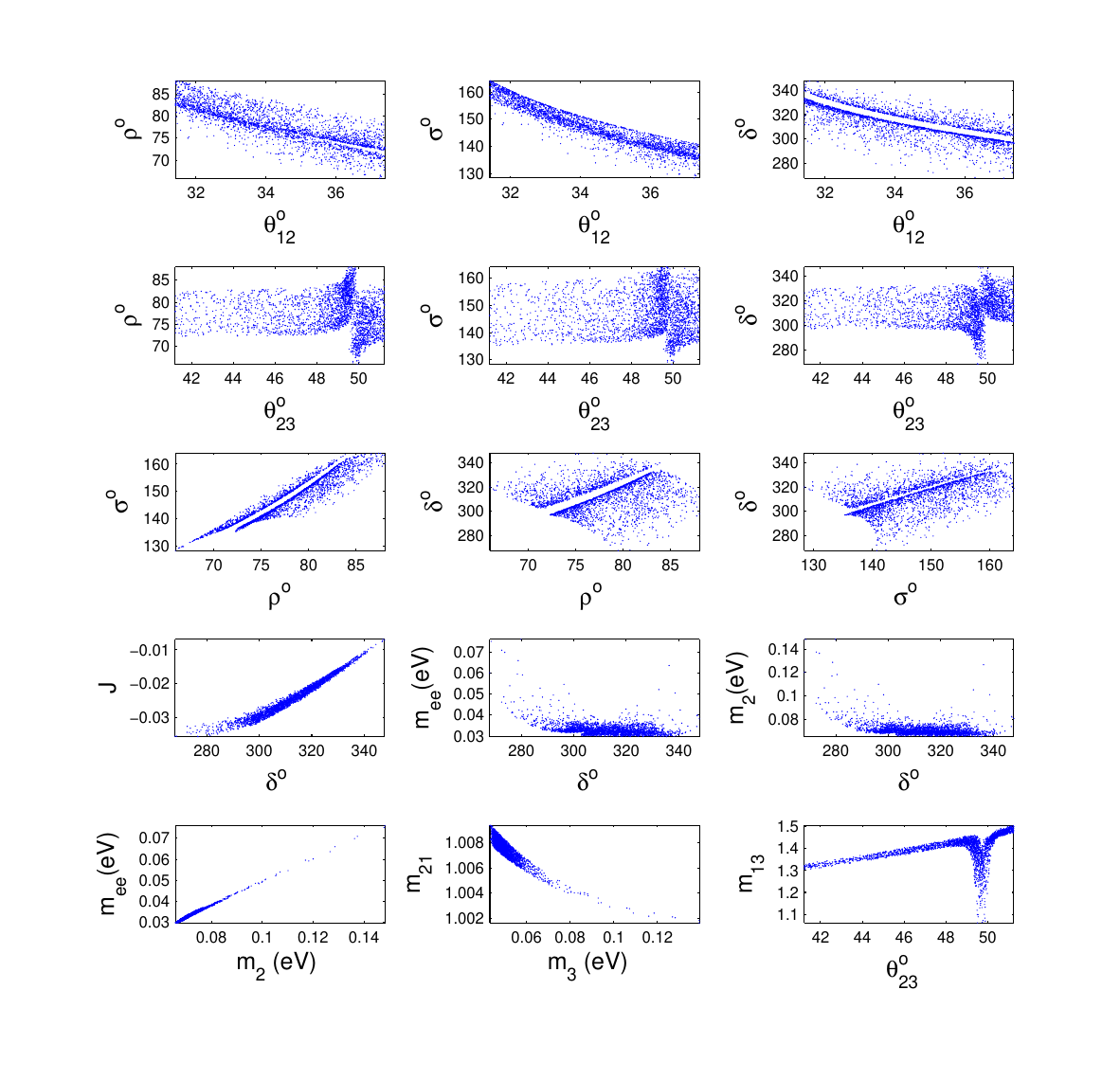}
\caption{The correlation plots for Texture III ($M_{11}+M_{22}=0$ \& $M_{12}-M_{33}=0$) in the case of inverted hierarchy,  at vanishing unphysical phases slice. The first and second rows represent the correlations between the mixing angles $(\theta_{12},\theta_{23})$ and the CP-violating phases. The third and fourth rows show the correlations amidst the CP-violating phases and the correlations between the Dirac phase $\delta$ and each of $J$, $m_{ee}$ and $m_2$ parameters respectively. The last row shows the degree of mass hierarchy plus the ($m_{ee}, m_2$) correlation.}
\label{IIIinv}
\end{figure}

\item {\bf Texture III: normal Hierarchy}

As stated before, this texture supports, for a very narrow band for $\t_y$ in $[49.26^o,50.06^o]$, the {\bf NH}. The correlation plots are shown in Fig. (\ref{IIInor}).

We reconstruct the neutrino mass matrix for a representative point which, for normal ordering, is taken as follows (the parameters $\t, \d$ are chosen to be near their best fit values):
\begin{equation}
\begin{aligned}
(\theta_{12},\theta_{23},\theta_{13})=&(35.38^{\circ},49.51^{\circ},8.53^{\circ}),\\
(\delta,\rho,\sigma)=&(190.46^{\circ},87.17^{\circ},99.18^{\circ}),\\
(m_{1},m_{2},m_{3})=&(0.0225\textrm{ eV},0.0241\textrm{ eV},0.0546\textrm{ eV}),\\
(m_{ee},m_{e})=&(0.0209\textrm{ eV},0.0242\textrm{ eV}).
\end{aligned}
\end{equation}
There is now a constructive interference between the contributions of $m_1$ and $m_2$ in determining $m_{ee}$ due to $\r \approx \s$. The corresponding neutrino mass matrix (in eV) is
\begin{equation}
M_{\nu}=\left( \begin {array}{ccc} -0.0208-0.0010i &  0.0093 + 0.0030i &  0.0065 - 0.0033i\\ \noalign{\medskip} 0.0093 + 0.0030i  &   0.0208+0.0010i &  0.0374-0.0019i
\\ \noalign{\medskip} 0.0065 - 0.0033i &  0.0374-0.0019i &  0.0093+0.0030i\end {array} \right).
\end{equation}

\begin{figure}[hbtp]
\hspace*{-3.5cm}
\includegraphics[width=22cm, height=16cm]{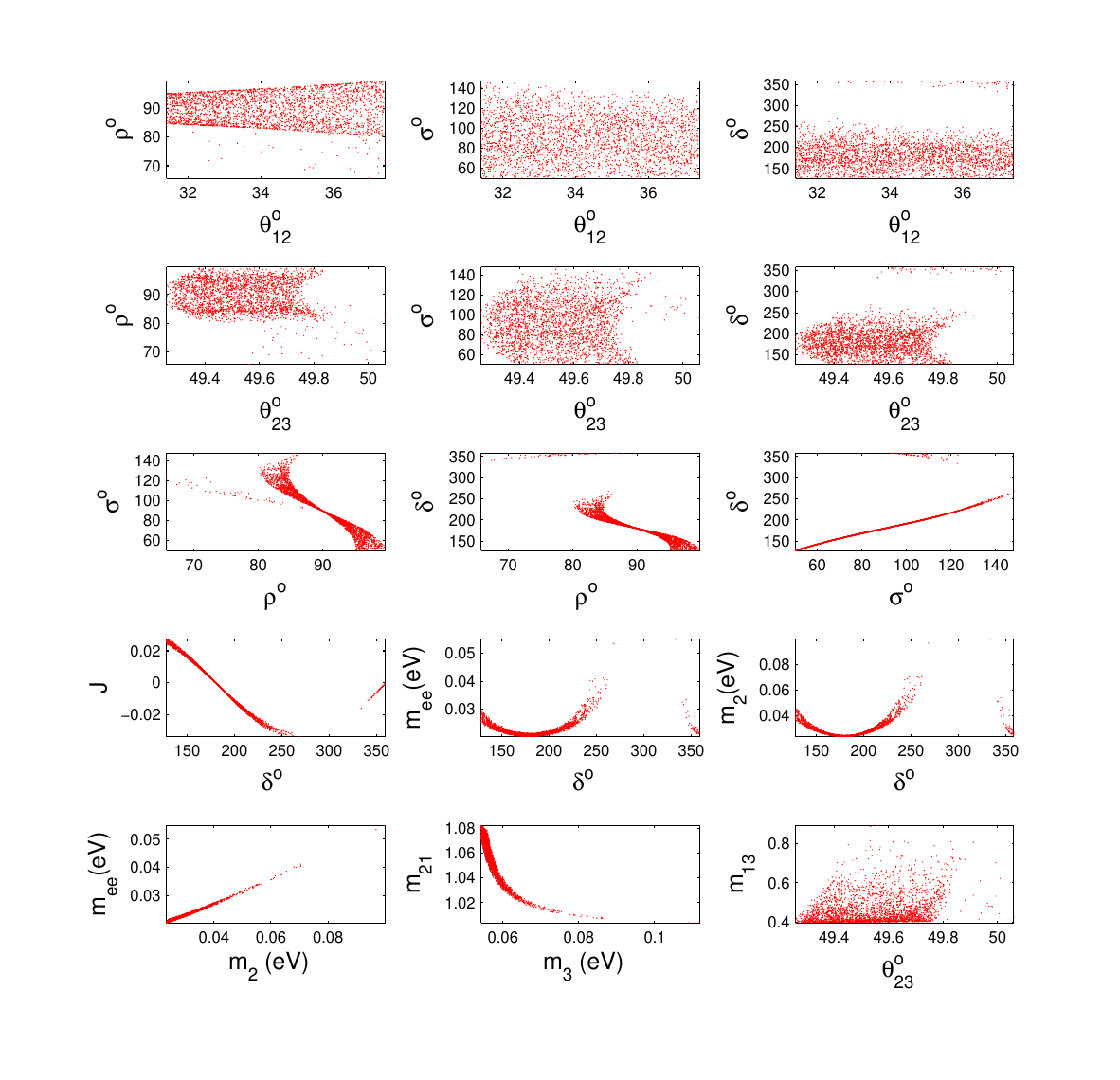}
\caption{The correlation plots for Texture III ($M_{11}+M_{22}=0$ \& $M_{12}-M_{33}=0$) in the case of normal hierarchy, at vanishing unphysical phases slice. The first and second rows represent the correlations between the mixing angles $(\theta_{12},\theta_{23})$ and the CP-violating phases. The third and fourth rows show the correlations amidst the CP-violating phases and the correlations between the Dirac phase $\delta$ and each of $J$, $m_{ee}$ and $m_2$ parameters respectively. The last row shows the degree of mass hierarchy plus the ($m_{ee}, m_2$) correlation.}
\label{IIInor}
\end{figure}

\end{enumerate}

\subsection{Numerical Results for general unphysical phases}
As said before, by including the unphysical phases, the neutrino mass matrix is described by $12$-parameters: ($m_1, m_2, m_3, \t_x, \t_y, \t_z, \d, \r, \s$ and $\phi_1, \phi_2, \phi_3$). By allowing for non vanishing unphysical phases, new solutions, respecting the mathematical constraint defining the texture and the experimental constraints, may be found. Whereas we studied in the previous subsection the case of vanishing unphysical phases, via adopting (`Specific' def.), and could, somehow, interpret the numerical results analytically, allowing now for non-vanishing unphysical phases, through (`Generalized' def.), gives results for the mass ratios turning out to be too complicated and lengthy to be displayed. We repeat that both (`Generalized' def.) and (`Specific' def.) are insensitive to unphysical phases, but in general they correspond to different textures and thus it is normal that they lead to different phenomenologies with correlations of (`Generalized' def.) containing those of (`Specific' def.).

We find that the three textures are viable accommodating both {\bf IH} \& {\bf NH} orderings, with predictions given in Table \ref{Predictions-nonVanishing}.
\begin{landscape}
\begin{table}[h]
\begin{center}
\begin{tabular} {c||c|c||c|c||c|c}
\hline
\hline
Observable & \multicolumn{2} {c}  {Pattern I} &
\multicolumn{2}  {c} {Pattern II} &
\multicolumn{2} {c} {Pattern III} \\
&\multicolumn{2} {c} {$M_{\n 22}=-M_{\n 33} \&  M_{\n 11}=+M_{\n 23}$} & \multicolumn{2} {c} {$M_{\n 11}=-M_{\n 33} \&  M_{\n 22}=+M_{\n 13}$} & \multicolumn{2} {c} {$M_{\n 11}=-M_{\n 22} \&  M_{\n 33}=+M_{\n 12}$} \\
\hline
\hline
Hierarchy & {\bf IH} & {\bf NH} & {\bf IH} & {\bf NH} & {\bf IH} &  {\bf NH}  \\
\hline
$\theta_{12}\equiv \t_x$ ($^\circ$) & $31.4 - 37.40$ & $31.4-37.40$ & $31.4-37.4$ & $31.4-37.4$  & $31.4 - 37.40$ & $31.4-37.40$   \\
\hline
$\theta_{23} \equiv \t_y$ ($^\circ$) &$41.16 - 51.25$ & $41.20-51.33$ & $41.16 - 51.25$ & $41.20-51.33$ &$41.16 - 51.25$ & $41.20-51.33$\\
\hline
$\theta_{13}\equiv \t_z$ ($^\circ$) &$8.17-8.96$ & $8.13-8.92$ & $8.17-8.96$ & $8.13-8.92$  &$8.17-8.96$ & $8.13-8.92$\\
\hline
$\delta$ ($^\circ$) & $200-353$ & $128-359$ & $200-353$  & $128-359 $ & $200-353$ & $128-359$ \\
\hline
$\rho$ ($^\circ$) & $0.03-179.99$ & $0-180$ & $0-180$ & $0-180$ &$0- 180$ & $0- 180$\\
\hline
$\sigma$ ($^\circ$) &$0.04 - 179.97 $ &$0- 180$ & $0- 180$ & $0-180$ &$0- 180$ & $0- 180$ \\
\hline
$\phi_1$ ($^\circ$) & $0.35 - 359.94 $ & $0.05-359.89$ & $0.03 - 359.93$ & $0.18-359.90$ &$0.07-359.81$ & $0.18-359.91$ \\
\hline
$\phi_2$ ($^\circ$) &$0.34 - 359.96 $ & $0.03-359.98$ & $0.06 - 359.97$ & $0.06-359.74$ & $0.07-359.98$ & $0.06-359.76$\\
\hline
$\phi_3$ ($^\circ$) &$0.55 - 359.85 $ & $0.07-359.87$ & $0.60 - 359.71$ & $0.14-359.97$ & $0.07-359.86$ & $0.00-359.96$\\
\hline
$m_1{\mbox{(eV)}}$ &$0.051-0.1755$ & $0.0156-0.1411$ & $0.0531-0.1631$ & $0.0237-0.1357$  &$0.0527-0.1285$ & $0.0219-0.0950$\\
\hline
$m_2{\mbox{(eV)}}$ &$0.0508-0.1757$ & $0.0180-0.1414$ & $0.0538-0.1633$ &$0.0252-0.1359$ &$0.0535-0.1287$ &$0.0235-0.0954$\\
\hline
$m_3{\mbox{(eV)}}$ &$0.0109-0.1685$ & $0.0521-0.1498$ & $0.0207-0.1553$ & $0.0551-0.1447$ &$0.0203-0.1184$ &$0.0544-0.1076$\\
\hline
$m_e{\mbox{(eV)}}$ &$0.0498-0.1754$ & $0.0180-0.1414$ & $0.0528-0.1630$ & $0.0253-0.136$ &$0.0524-0.1283$ & $0.0236-0.0955$\\
\hline
$m_{ee}{\mbox{(eV)}}$ &$0.0127-0.1682$ & $0.0163-0.1353$ & $0.0311-0.1031$ & $0.0138-0.0796$ &$0.0293-0.0769$ & $0.0147-0.0538$\\
\hline
$\Sigma {\mbox{(eV)}}$ & $0.1126-0.5196$ & $0.0857-0.4323$ & $0.1280-0.4817$ & $0.1042-0.4163$ & $0.1266-0.3756$ & $0.0998-0.2980$\\
\hline
\hline
 \end{tabular}
 \end{center}
 \caption{The various predictions for the ranges of the neutrino measurable parameters for three viable $S_4$-inspired textures at 3-$\sigma$ level.}
\label{Predictions-nonVanishing}
 \end{table}
\end{landscape}
Compared to the vanishing unphysical phases correlations, the new ones, when one includes the unphysical phases, get ``diluted" and quite often disappear completely when lacking the characteristic features distinguishing the vanishing case. There are few generic features in the new correlations when compared to the vanishing case, which can be summarized as follows.
\begin{itemize}
\item The pairwise correlations between the phases ($\d, \r, \s$) and the mixing angles ($\t_x \equiv \t_{12}, \t_y\equiv \t_{23}$) disappear.

\item There are still persistent pairwise correlations amidst the phases  ($\d, \r, \s$), which are different from the vanishing unphysical phases  case ($\phi_k=0, k=1,2,3$), as turning on the unphysical phases ($\phi_k$) makes all the admissible ranges for ($\d, \r, \s$) viable. These correlations take the form of bands.

\item The correlations of $J$ versus $\d$ follow the sinusoidal curve $J \propto \sin \d$, as it should be.

\item The correlations ($m_{ee}, \d$), ($m_2, \d$) and ($m_{13}, \t_y \equiv \t_{23}$) disappear.

\item  The pairwise correlations ($m_{ee}, m_2$) and ($m_{21}, m_3$) persist with the same shape as in the vanishing case, where $m_{ee}$ increases with increasing $m_2$, while $m_{21}$ decreases with increasing $m_3$.

\end{itemize}

Finally, we see that adopting the cosmological most stringent bound (Eq. \ref{cosTB}) will rule out all the patterns, except the pattern I in normal ordering since the lowest bound $0.0857~\text{eV}$  lies below the tough bound. For this tough bound we get the corresponding acceptable ranges for the measurable neutrino parameters in Table (\ref{predictionToughest}):

\begin{table}[h]
\hspace{-2.5cm}
\footnotesize
\begin{tabular} {||c|c|c|c|c|c|c|c||}
\hline
$\t_{12}(^{o})$ & $\t_{23}(^{o})$ & $\t_{13}(^{o})$ & $\d(^{o})$ & $\r(^{o})$ & $\s(^{o})$ & $\phi_1(^{o})$ & $\phi_2(^{o})$   \\
\hline
$31.40-37.39$ & $43.82-46.25$ & $8.13-8.92$ &  $128.34-358.80$ &  $0.05-179.82$ &  $0.08-179.88$ &  $0.05-359.92$ & $0.99-359.51$ \\
\hline \hline
 $\phi_3(^{o})$  & $m_1~\text{(eV)}$ & $m_2~\text{(eV)}$ & $m_3~\text{(eV)}$  & $m_e~\text{(eV)}$ & $m_{ee}~\text{(eV)}$ & $\Sigma~\text{(eV)}$ \\
\hline
    $0.76-359.22$ & $0.0155-0.0176$ & $0.0178-0.0197$ & $0.0521-0.0540$ & $0.0179-0.0198$ & $0.0163-0.0189$ & $0.0855-0.09$ \\
    \hline \hline
\end{tabular}
 \caption{The various predictions for the ranges of the neutrino measurable parameters for the only surviving pattern I with Normal ordering, when adopting the toughest cosmological bound ($\Sigma < 0.09 ~\text{eV}$).}
\label{predictionToughest}
 \end{table}

\subsubsection{Unphysical phases correlations}
Although correlations involving unphysical phases are somehow `trivial', consisting of a union of straight lines parallel to the unphysical phase axis, according to (`Specific' def.) or (`Generalized' def.), since all values for the unphysical phases are accepted once a point is accepted, however, and for completeness purposes, we shall start this subsection involving non-vanishing unphysical phases by noting that within the (`Mathematical' def.), correlations involving the unphysical phases, albeit devoid of physical relevance, are bound to exist in order to meet the mathematical constraint characterizing the texture. Let's clarify this point by assuming the Texture I ($M_{22}+M_{33}=0$ \& $M_{11}-M_{23}=0$). Then using Eq. (\ref{massmatrixelementsphase}) we have
\bea
e^{i2\phi_2} M_{22}(\phi_k=0) + e^{i2\phi_3} M_{33}(\phi_k=0) = 0 &,&
e^{i2\phi_1} M_{11}(\phi_k=0) - e^{i(\phi_2+\phi_3)} M_{23}(\phi_k=0) = 0 \nn \\
\Rightarrow
e^{i(\phi_{21}-3\phi_{31})} &=& \!\!\! - \frac{M_{33}(\phi_k=0)  M_{23}(\phi_k=0)}{M_{11}(\phi_k=0) M_{23}(\phi_k=0)}: \phi_{kj}=\phi_k -\phi_j
\eea
This relation would impose certain correlations between the unphysical phases ($\phi_{21}$ and $\phi_{31}$). Since $\r,\s$ are functions of ($\t_{x(y,z)}, \phi_k, \d$), then it is possible to get correlations among Majorana phases ($\r, \s$) and the unphysical phases. We have checked this fact, and could generate the corresponding correlations involving the unphysical phases, bearing in mind that they correspond to (`Mathematical' def.) and thus are of no physical significance.

Having shown why the unphysical phases do have correlations with other parameters, within (`Mathematical' def.), and taking into account the equivalence of (`Mathemdatical' def.) and (`Generalized' def.) in relation to the physical correlations, we move on now to study the three textures with non-vanishing unphysical phases, where the ``clear'' stated correlations are divided into three lines each containing three correlations: the first line includes those amidst CP-phases ($\r, \s, \d$), the second lines includes mass correlations involving ($m_3, m_2, m_{21}, m_{ee}$) and the correlation ($\d, J$), whereas the last line, corresponding to (`mathematical' def.) and put in just for verfication purposes, includes the correlations of the unphysical phase ($\phi_{31}$) with the Majorana phases ($\r, \s$) and with the unphysical phase ($\phi_{12}$).

\subsubsection{Texture I ($M_{22}+M_{33}=0$ \& $M_{11}-M_{23}=0$) with non-vanishing unphysical phases}
Unlike the vanishing unphysical phases case, where only {\bf IH} ordering was viable,  allowing now for non-vanishing unphysical phases makes the two orderings viable for this texture.

\begin{enumerate}

\item {\bf Inverted ordering}

The correlation plots are shown in Fig. (\ref{Iinv-nvanishing}).

We reconstruct the neutrino mass matrix for a representative point which, for inverted ordering, is taken as follows (the parameters $\t, \d$ are chosen near their best fit values):
\begin{equation}
\begin{aligned}
(\theta_{12},\theta_{23},\theta_{13})=&(35.38^{\circ},49.44^{\circ},8.55^{\circ}),\\
(\delta,\rho,\sigma)=&(289.77^{\circ},64.90^{\circ},5.78^{\circ}),\\
(\phi_1,\phi_2,\phi_3)=&(234.95^{\circ},290.46^{\circ},274.34^{\circ}),\\
(m_{1},m_{2},m_{3})=&(0.0589\textrm{ eV},0.0595\textrm{ eV},0.0326\textrm{ eV}),\\
(m_{ee},m_{e}, \Sigma)=&(0.0337\textrm{ eV},0.0337\textrm{ eV}, 0.1511\textrm{ eV}).
\end{aligned}
\end{equation}
The corresponding neutrino mass matrix (in eV) is
\begin{equation}
M_{\nu}=\left( \begin {array}{ccc} -0.0298-0.0157i &  -0.0280-0.0136i &  0.0323 + 0.0172i\\ \noalign{\medskip} -0.0280-0.0136i  &   -0.0030+0.0047i &  -0.0298-0.0157i
\\ \noalign{\medskip} 0.0323 + 0.0172i &   -0.0298-0.0157i &  0.0030-0.0047i\end {array} \right).
\end{equation}

\begin{figure}[hbtp]
\hspace*{-3.5cm}
\includegraphics[width=22cm, height=16cm]{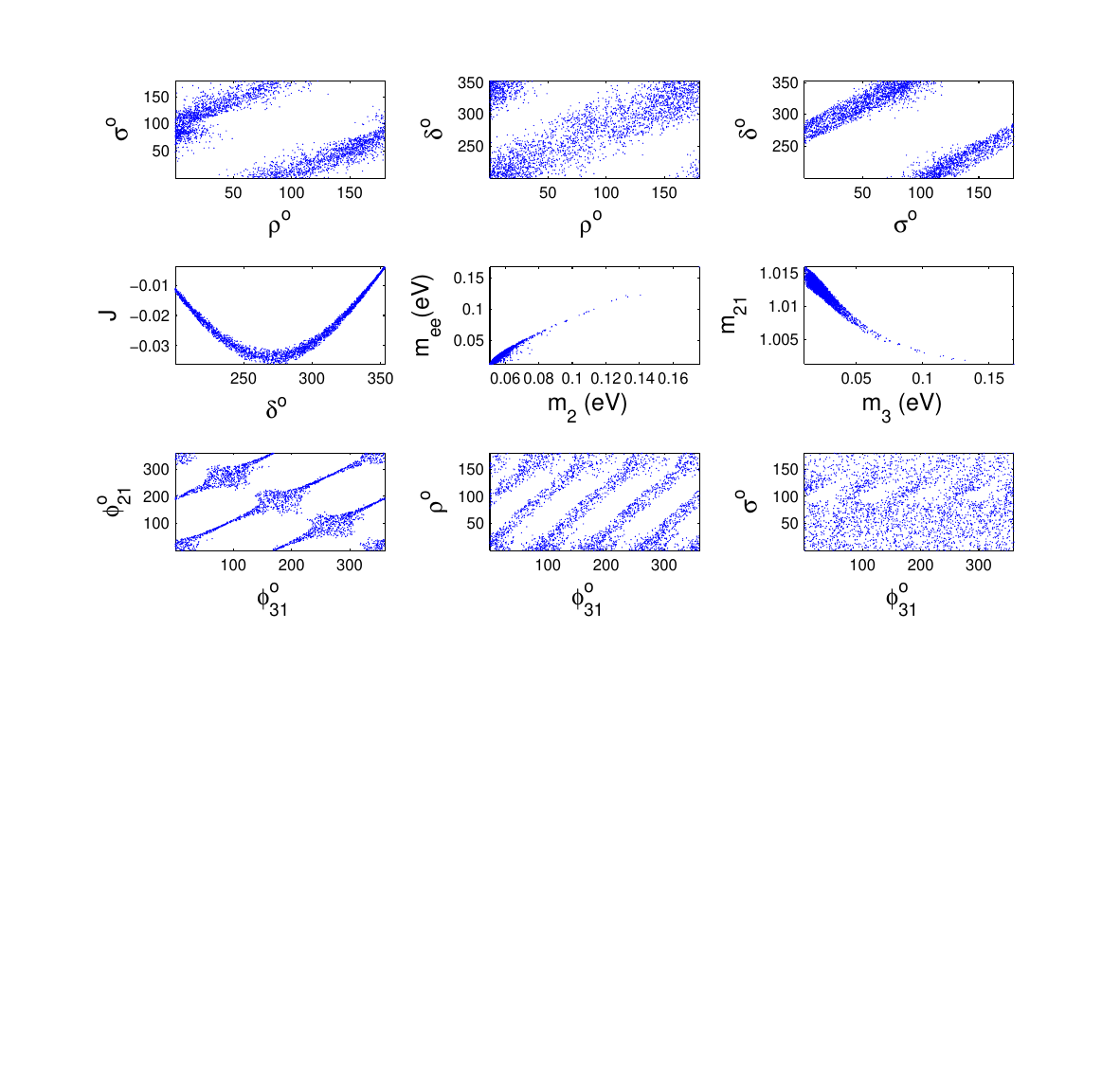}
\vspace*{-7.5cm}
\caption{The non-disappearing correlation plots for Texture I ($M_{22}+M_{33}=0$ \& $M_{11}-M_{23}=0$) in the case of inverted hierarchy, with non vanishing unphysical phases.}
\label{Iinv-nvanishing}
\end{figure}

\item {\bf Normal ordering}

The correlation plots are shown in Fig. (\ref{Inor-nvanishing}).

We reconstruct the neutrino mass matrix for a representative point which, for normal ordering, is taken as follows (the parameters $\t, \d$ are chosen near their best fit values):
\begin{equation}
\begin{aligned}
(\theta_{12},\theta_{23},\theta_{13})=&(34.71^{\circ},44.44^{\circ},8.52^{\circ}),\\
(\delta,\rho,\sigma)=&(192.67^{\circ},154.93^{\circ},41.44^{\circ}),\\
(\phi_1,\phi_2,\phi_3)=&(141.71^{\circ},359.82^{\circ},277.55^{\circ}),\\
(m_{1},m_{2},m_{3})=&(0.0444\textrm{ eV},0.0453\textrm{ eV},0.0667\textrm{ eV}),\\
(m_{ee},m_{e}, \Sigma)=&(0.0236\textrm{ eV},0.0453\textrm{ eV}, 0.1563\textrm{ eV}).
\end{aligned}
\end{equation}
The corresponding neutrino mass matrix (in eV) is
\begin{equation}
M_{\nu}=\left( \begin {array}{ccc} -0.0029-0.0234i &  0.0114+0.0245i &  -0.022 + 0.0168i\\ \noalign{\medskip} 0.0114+0.0245i  &   0.0421+0.0114i &  -0.0029-0.0234i
\\ \noalign{\medskip} -0.022 + 0.0168i &  -0.0029-0.0234i &  -0.0421-0.0114i\end {array} \right).
\end{equation}

\begin{figure}[hbtp]
\hspace*{-3.5cm}
\includegraphics[width=22cm, height=16cm]{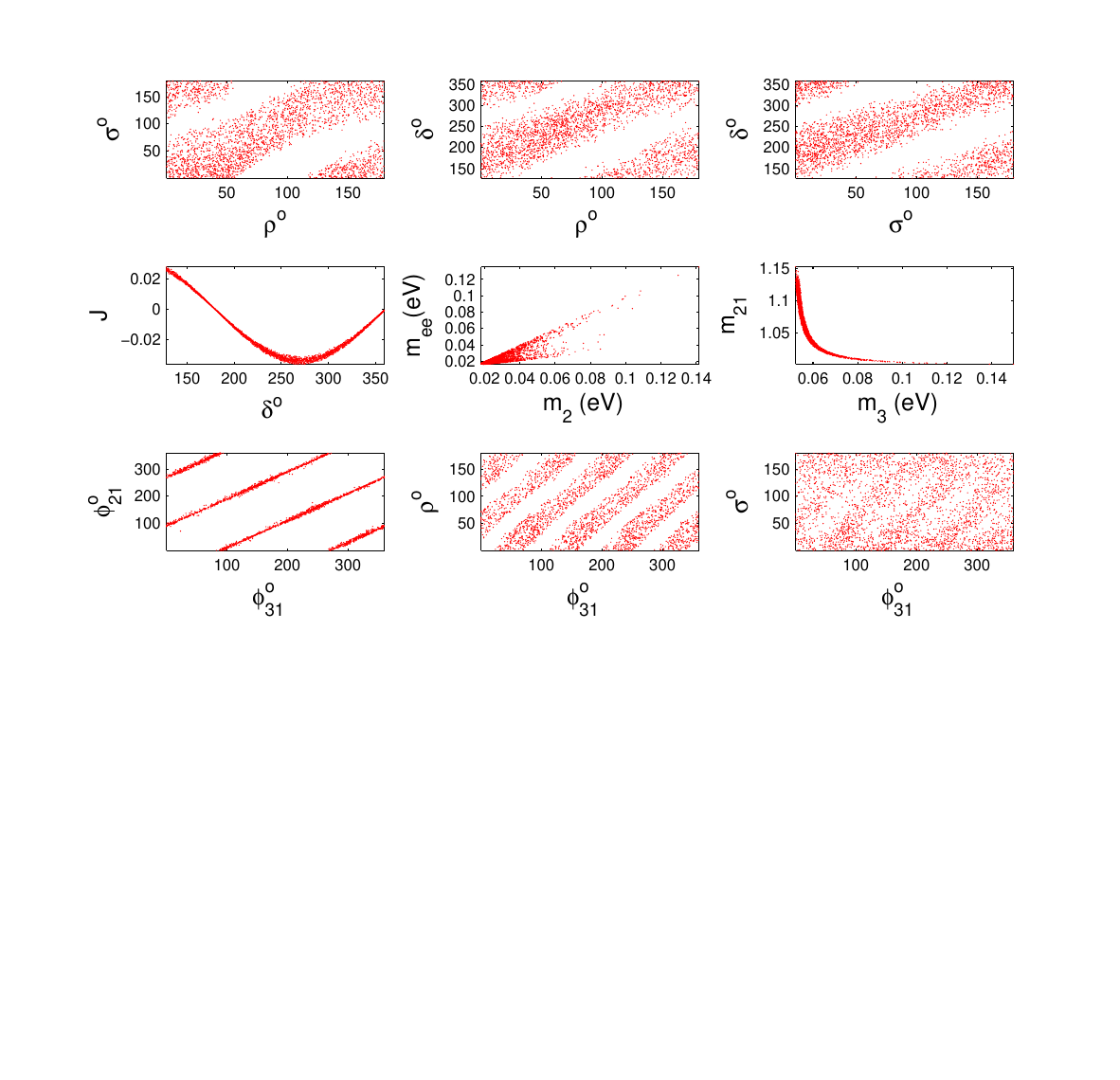}
\vspace*{-7.5cm}
\caption{The non-disappearing correlation plots for Texture I ($M_{22}+M_{33}=0$ \& $M_{11}-M_{23}=0$) in the case of normal hierarchy, with non vanishing unphysical phases.}
\label{Inor-nvanishing}
\end{figure}

\item {\bf Normal ordering at most stringent cosmological bound ($\Sigma < 0.09$ eV)}

As is seen in Table (\ref{predictionToughest}), there is a slightly small parameter space surviving when adopting the cosmological toughest bound of (Eq. \ref{cosTB}). The correlation plots are shown in Fig. (\ref{Inor-nvanishing-toughest}). Compared with the correlation plots corresponding to the less stringent bound (Fig \ref{Inor-nvanishing}), which would represent a superset to plots of (Fig. \ref{Inor-nvanishing-toughest}), we see that both sets of plots are similar, except for plots ($(m_{ee}, m_2)$ and $(m_{21},m_3)$) due to the very restricted mass range allowed in the cosmological tough bound case.

In this case, we reconstruct the neutrino mass matrix for the following representative point ($\t, \d$ are chosen near their best fit values):
\begin{equation}
\begin{aligned}
(\theta_{12},\theta_{23},\theta_{13})=&(31.82^{\circ},44.65^{\circ},8.6^{\circ}),\\
(\delta,\rho,\sigma)=&(195.99^{\circ},174.09^{\circ},11.17^{\circ}),\\
(\phi_1,\phi_2,\phi_3)=&(254.56^{\circ},293.73^{\circ},204.78^{\circ}),\\
(m_{1},m_{2},m_{3})=&(0.0169\textrm{ eV},0.0190\textrm{ eV},0.0526\textrm{ eV}),\\
(m_{ee},m_{e}, \Sigma)=&(0.0177\textrm{ eV},0.0190\textrm{ eV}, 0.0886\textrm{ eV}).
\end{aligned}
\end{equation}
The corresponding neutrino mass matrix (in eV) is
\begin{equation}
M_{\nu}=\left( \begin {array}{ccc} -0.0149+0.0095i &  -0.0029+0.0027i &  -0.0039 + 0.0044i\\ \noalign{\medskip}-0.0029+0.0027i  &   -0.0249-0.0239i &  -0.0149+0.0095i
\\ \noalign{\medskip} -0.0039 + 0.0044i &  -0.0149+0.0095i &  0.0249+0.0239i\end {array} \right).
\end{equation}

\begin{figure}[hbtp]
\hspace*{-3.5cm}
\includegraphics[width=22cm, height=16cm]{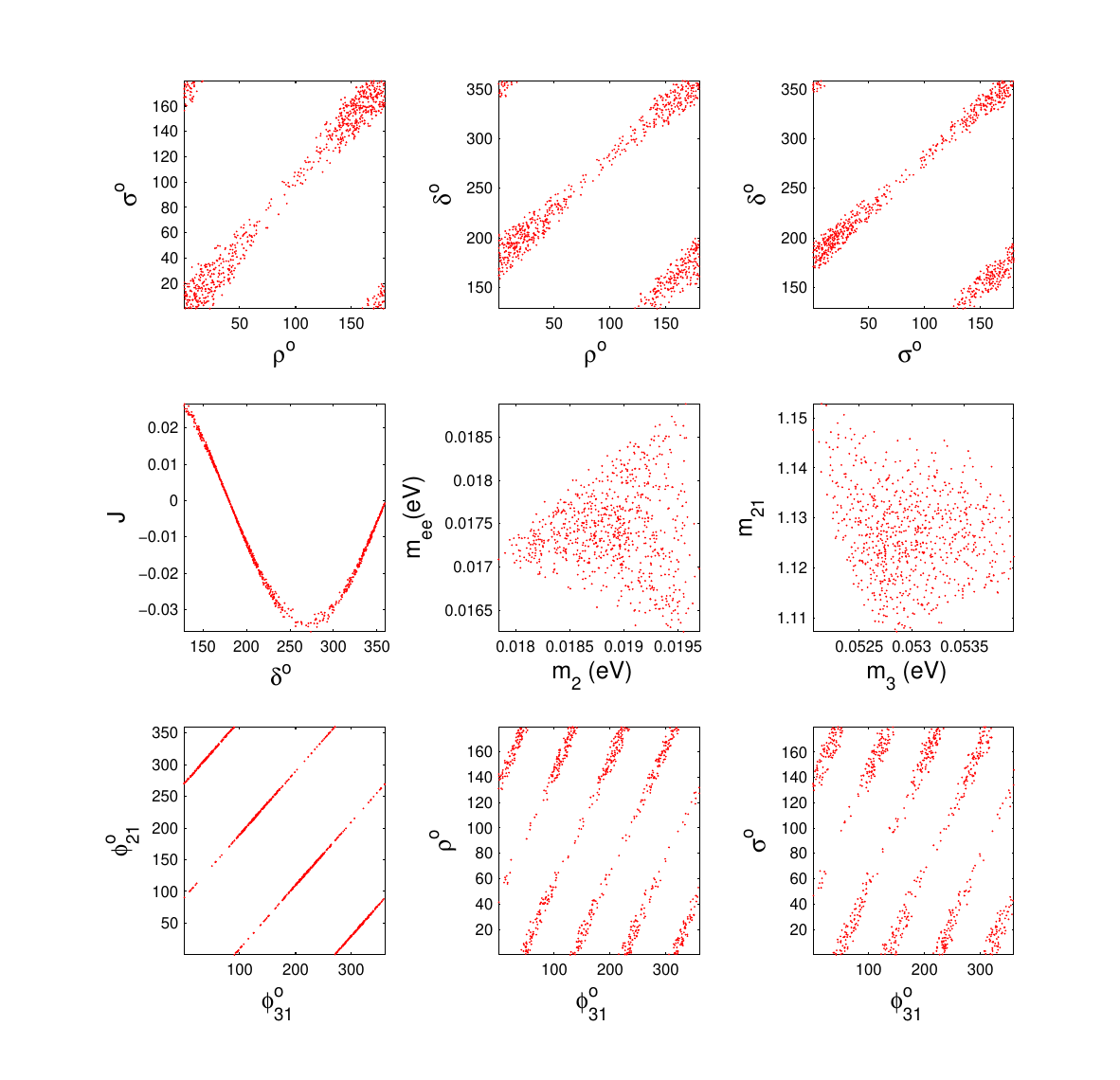}
\caption{The non-disappearing correlation plots for Texture I ($M_{22}+M_{33}=0$ \& $M_{11}-M_{23}=0$) in the case of normal hierarchy, with non vanishing unphysical phases, adopting the toughest cosmological bound ($\Sigma < 0.09~\text{e.V.}$).}
\label{Inor-nvanishing-toughest}
\end{figure}

\end{enumerate}

\subsubsection{Texture II ($M_{11}+M_{33}=0$ \& $M_{22}-M_{13}=0$) with non-vanishing unphysical phases}
Unlike the vanishing unphysical case, where only {\bf IH} ordering was viable,  allowing now for non-vanishing unphysical phases makes the two orderings viable for this texture.

\begin{enumerate}

\item {\bf Inverted ordering}

The correlation plots are shown in Fig. (\ref{IIinv-nvanishing}).

We reconstruct the neutrino mass matrix for a representative point which, for inverted ordering, is taken as follows (the parameters $\t, \d$ are chosen near their best fit values):
\begin{equation}
\begin{aligned}
(\theta_{12},\theta_{23},\theta_{13})=&(34.42^{\circ},48.32^{\circ},8.55^{\circ}),\\
(\delta,\rho,\sigma)=&(287.95^{\circ},63.19^{\circ},113.72^{\circ}),\\
(\phi_1,\phi_2,\phi_3)=&(60.35^{\circ},227.85^{\circ},232.85^{\circ}),\\
(m_{1},m_{2},m_{3})=&(0.0540\textrm{ eV},0.0546\textrm{ eV},0.0232\textrm{ eV}),\\
(m_{ee},m_{e}, \Sigma)=&(0.0362\textrm{ eV},0.0537\textrm{ eV}, 0.1318\textrm{ eV}).
\end{aligned}
\end{equation}
The corresponding neutrino mass matrix (in eV) is
\begin{equation}
M_{\nu}=\left( \begin {array}{ccc} 0.0025-0.0361i &  -0.0012-0.0322i &  -0.0003 + 0.0231i\\ \noalign{\medskip} -0.0012-0.0322i  &   -0.0003+0.0231i &  -0.0035-0.0074i
\\ \noalign{\medskip}-0.0003 + 0.0231i &   -0.0035-0.0074i &  -0.0025+0.0361i\end {array} \right).
\end{equation}

\begin{figure}[hbtp]
\hspace*{-3.5cm}
\includegraphics[width=22cm, height=16cm]{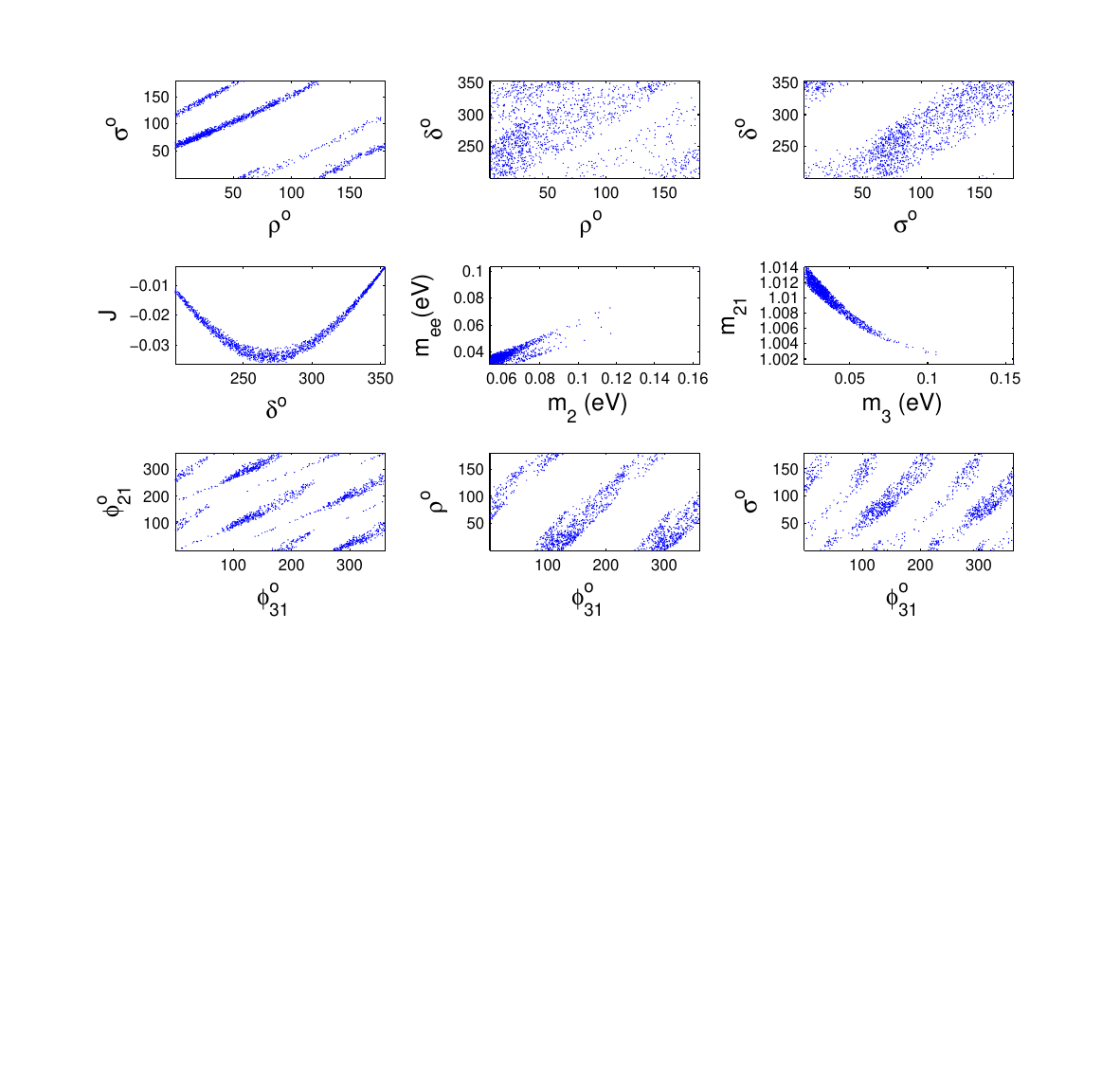}
\vspace*{-7.5cm}
\caption{The non-disappearing correlation plots for Texture II ($M_{11}+M_{33}=0$ \& $M_{22}-M_{13}=0$) in the case of inverted hierarchy, with non vanishing unphysical phases.}
\label{IIinv-nvanishing}
\end{figure}

\item {\bf Normal ordering}

The correlation plots are shown in Fig. (\ref{IInor-nvanishing}).

We reconstruct the neutrino mass matrix for a representative point which, for normal ordering, is taken as follows (the parameters $\t, \d$ are chosen near their best fit values):
\begin{equation}
\begin{aligned}
(\theta_{12},\theta_{23},\theta_{13})=&(35.47^{\circ},48.87^{\circ},8.59^{\circ}),\\
(\delta,\rho,\sigma)=&(196.29^{\circ},61.18^{\circ},130.38^{\circ}),\\
(\phi_1,\phi_2,\phi_3)=&(247.78^{\circ},207.92^{\circ},229.96^{\circ}),\\
(m_{1},m_{2},m_{3})=&(0.0408\textrm{ eV},0.0418\textrm{ eV},0.0648\textrm{ eV}),\\
(m_{ee},m_{e}, \Sigma)=&(0.0173\textrm{ eV},0.0418\textrm{ eV}, 0.1474\textrm{ eV}).
\end{aligned}
\end{equation}
The corresponding neutrino mass matrix (in eV) is
\begin{equation}
M_{\nu}=\left( \begin {array}{ccc} 0.0045-0.0167i &  -0.0236+0.0088i &  0.0221 + 0.0179i\\ \noalign{\medskip} -0.0236+0.0088i &   0.0221+0.0179i &  0.0043+0.0404i
\\ \noalign{\medskip} 0.0221 + 0.0179i &  0.0043+0.0404i &  -0.0045+0.0167i\end {array} \right).
\end{equation}

\begin{figure}[hbtp]
\hspace*{-3.5cm}
\includegraphics[width=22cm, height=16cm]{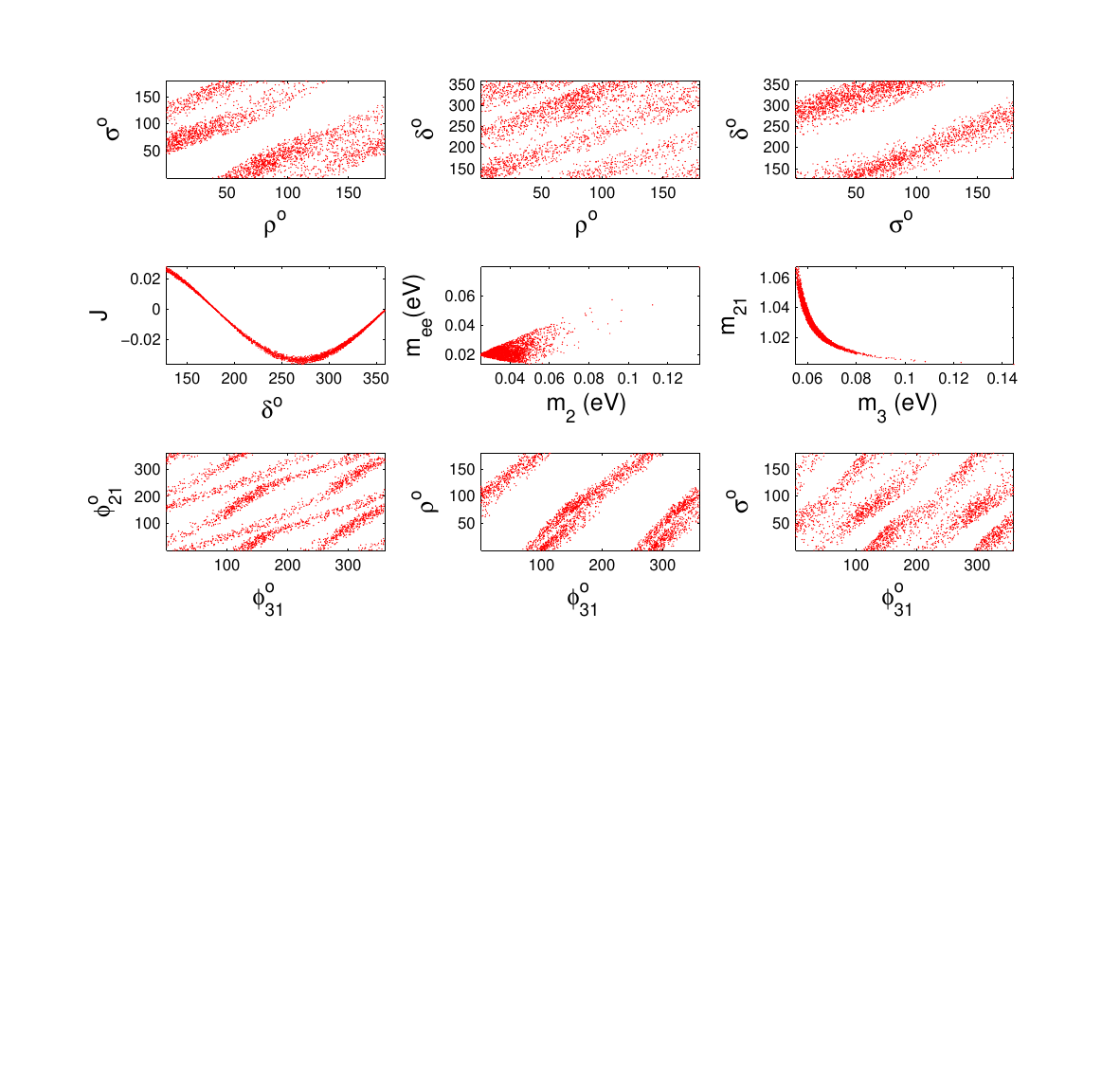}
\vspace*{-7.5cm}
\caption{The non-disappearing correlation plots for Texture II ($M_{11}+M_{33}=0$ \& $M_{22}-M_{13}=0$) in the case of normal hierarchy, with non vanishing unphysical phases.}
\label{IInor-nvanishing}
\end{figure}

\end{enumerate}

\subsubsection{Texture III ($M_{11}+M_{22}=0$ \& $M_{12}-M_{33}=0$)  with non-vanishing unphysical phases}
The vanishing unphysical phases case supported two  orderings {\bf IH \& NH}, but the {\bf NH} was concentrated at $\t_y \approx 50^o$. Allowing for non-vanishing unphysical phases enlarges this region for the {\bf NH}.

\begin{enumerate}

\item {\bf Inverted ordering}

The correlation plots are shown in Fig. (\ref{IIIinv-nvanishing}).

We reconstruct the neutrino mass matrix for a representative point which, for inverted ordering, is taken as follows (the parameters $\t, \d$ are chosen near their best fit values):
\begin{equation}
\begin{aligned}
(\theta_{12},\theta_{23},\theta_{13})=&(34.72^{\circ},48.35^{\circ},8.55^{\circ}),\\
(\delta,\rho,\sigma)=&(285.78^{\circ},124.79^{\circ},68.25^{\circ}),\\
(\phi_1,\phi_2,\phi_3)=&(26.08^{\circ},237.12^{\circ},55.90^{\circ}),\\
(m_{1},m_{2},m_{3})=&(0.0574\textrm{ eV},0.0580\textrm{ eV},0.0276\textrm{ eV}),\\
(m_{ee},m_{e}, \Sigma)=&(0.0346\textrm{ eV},0.0571\textrm{ eV}, 0.1430\textrm{ eV}).
\end{aligned}
\end{equation}
The corresponding neutrino mass matrix (in eV) is
\begin{equation}
M_{\nu}=\left( \begin {array}{ccc} 0.0021-0.0345i &  0.0066+0.0229i &  0.0046+0.0384i\\ \noalign{\medskip} 0.0066+0.0229i  &   -0.0021+0.0345i &  0.0119+0.0050i
\\ \noalign{\medskip} 0.0046+0.0384i &  0.0119+0.0050i &  0.0066+0.0229i\end {array} \right).
\end{equation}

\begin{figure}[hbtp]
\hspace*{-3.5cm}
\includegraphics[width=22cm, height=16cm]{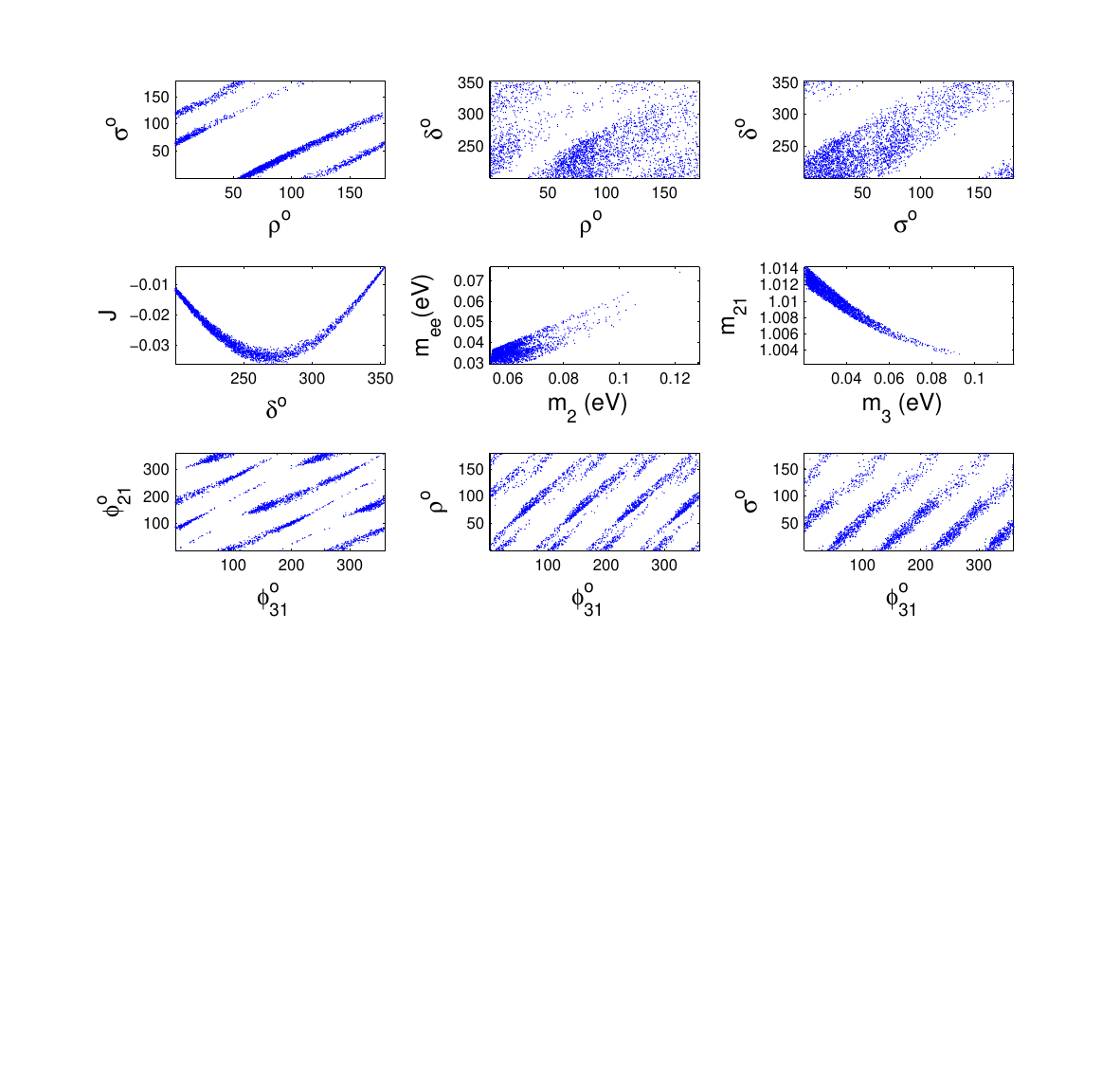}
\vspace*{-7.5cm}
\caption{The non-disappearing correlation plots for Texture III ($M_{11}+M_{22}=0$ \& $M_{12}-M_{33}=0$)  in the case of inverted hierarchy, with non vanishing unphysical phases.}
\label{IIIinv-nvanishing}
\end{figure}

\item {\bf Normal ordering}

The correlation plots are shown in Fig. (\ref{IIInor-nvanishing}).

We reconstruct the neutrino mass matrix for a representative point which, for normal ordering, is taken as follows (the parameters $\t, \d$ are chosen near their best fit values):
\begin{equation}
\begin{aligned}
(\theta_{12},\theta_{23},\theta_{13})=&(35.45^{\circ},48.51^{\circ},8.56^{\circ}),\\
(\delta,\rho,\sigma)=&(195.77^{\circ},81.82^{\circ},109.82^{\circ}),\\
(\phi_1,\phi_2,\phi_3)=&(111.95^{\circ},293.36^{\circ},32.54^{\circ}),\\
(m_{1},m_{2},m_{3})=&(0.0240\textrm{ eV},0.0255\textrm{ eV},0.0554\textrm{ eV}),\\
(m_{ee},m_{e}, \Sigma)=&(0.0202\textrm{ eV},0.0256\textrm{ eV}, 0.1050\textrm{ eV}).
\end{aligned}
\end{equation}
The corresponding neutrino mass matrix (in eV) is
\begin{equation}
M_{\nu}=\left( \begin {array}{ccc} 0.0139+0.0147i &  0.0015+0.0118i &  -0.0007+0.0103i\\ \noalign{\medskip} 0.0015+0.0118i &   -0.0139-0.0147i &  0.0303-0.0225i
\\ \noalign{\medskip} -0.0007+0.0103i &   0.0303-0.0225i &  0.0015+0.0118i\end {array} \right).
\end{equation}

\begin{figure}[hbtp]
\hspace*{-3.5cm}
\includegraphics[width=22cm, height=16cm]{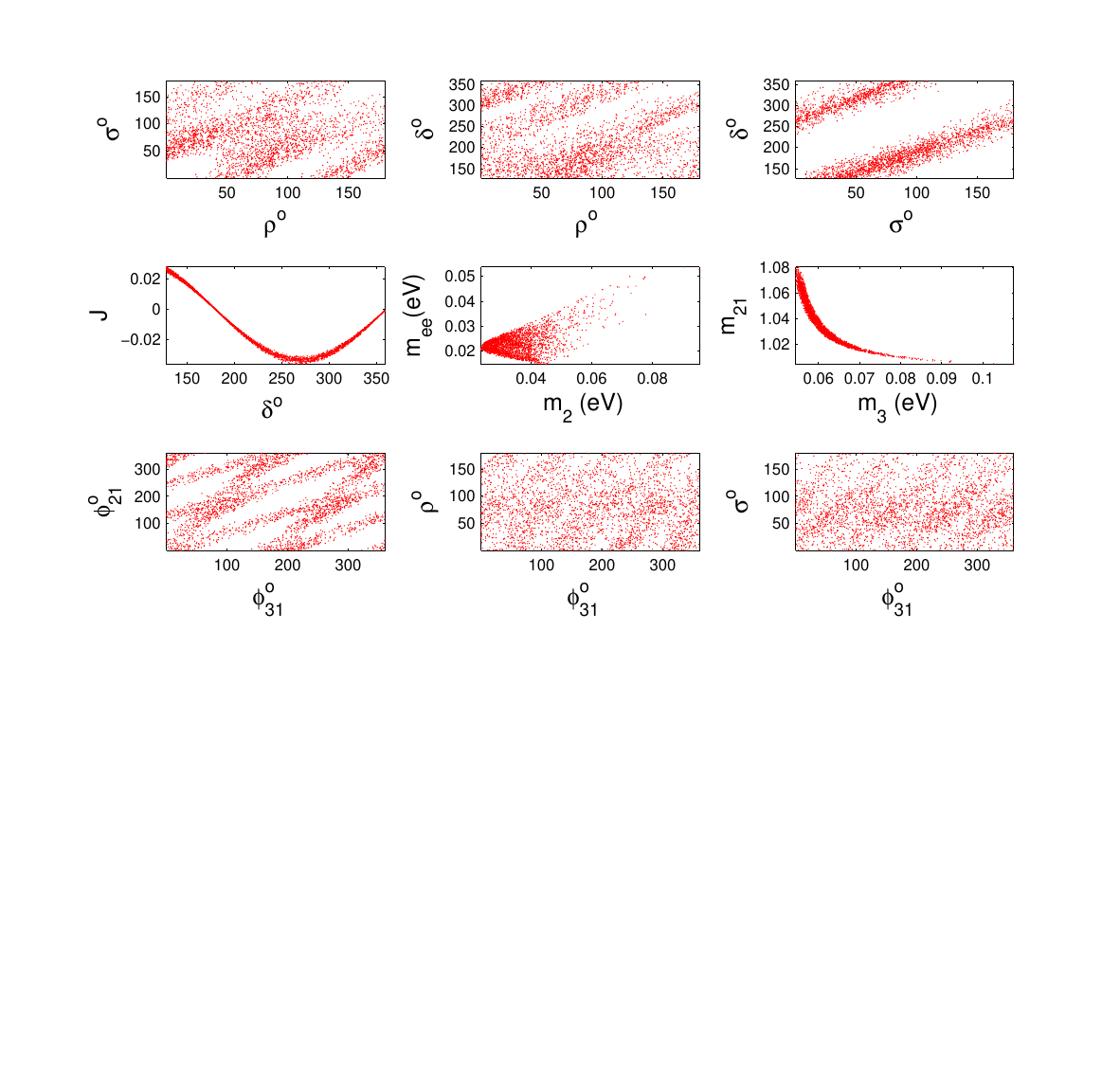}
\vspace*{-7.5cm}
\caption{The non-disappearing correlation plots for Texture III ($M_{11}+M_{22}=0$ \& $M_{12}-M_{33}=0$) in the case of normal hierarchy, with non vanishing unphysical phases.}
\label{IIInor-nvanishing}
\end{figure}

\end{enumerate}


\section{Theoretical realization}
 The symmetry is based on the non-abelian group $S_4$, and the matter content is extended to include new scalars. While
Abelian symmetries are simple and were used abundantly
within type-I and type-II seesaw scenarios (e.g., see
Refs. \cite{ismael_npb,ismael_2021} and references therein), non-Abelian discrete
symmetries are considered a far richer and more interesting
choice for the flavor sector. The group $S_4$ is one of the simplest choices of non-abelian discrete groups that contain two and three dimensional representations, in such a way that the three flavors are embedded in a non trivial way as a representation of the group.  This $S_4$ group  has been, actually, used extensively as the flavor symmetry group in model building, e.g. \cite{ismael_2022,thapa_2021,petcov_2017,scot_2013,lam_2008,merlo_2009,grimus_2009,zhang_2011,merlo_2013,king_2011}. We do not treat the question of the scalar potential and how to find its general form under the imposed symmetry which would give the required VEVs, as this goes beyond the scope of the paper. Having new scalars may lead to rich phenomenology at colliders, and requesting only one SM-like Higgs at low scale is not a trivial task, and requires generally fine tuning. We stress again that the realization presented here corresponds well to the (`Generalized' def.) rather than the (`Specific' def.) since the obtained mass matrix satisfying the characterizing mathematical constraint is not guaranteed to be corresponding to vanishing unphysical phases.

\subsection{$S_4$-non abelian group realization of the texture ($M_{\n33}=-M_{\n22}$  and $M_{\n11}=+M_{\n23}$ )}
\label{subsection-s4}
 We present now one realization of the texture I, which remains valid for the other two textures by applying transpositions on the indices: $1\leftrightarrow 2$ ($1\leftrightarrow 3$) to get the texture II (III). For completeness,  we summarize the irreducible representations (irreps) of $(S_4)$ in appendix (\ref{appendix-s4}), and state the corresponding multiplication rules in appendix (\ref{appendix_mult}).

\subsubsection{Type-II Seesaw Matter Content:}
\label{subsubsection-s4-matter}
We present a type-II seesaw scenario leading to a neutrino mass matrix of the required form. The matter content is summarized in Table (\ref{matter-content-S4})

 \begin{table}[h]
\caption{matter content and symmetry transformations, leading to texture I. $\Delta_j= \left( \begin{array} {cc} \Delta^+_j &  \Delta^{++}_ j \\  \Delta^{0}_ j & -\Delta^+_j  \end{array}\right)$, $j=1,\ldots,4$ with $i=1,2,3$ is a family index.}
\centering
\begin{tabular}{ccccccccc}
\hline
\hline
Fields & $D_{L_i}$ & $\Delta_i$ & $\Delta_4$ & $\ell_{R_i}$ & $\phi_I$ & $\phi_{II}$ & $\phi_{III}$ & $\phi_{III}'$   \\
\hline
\hline
$SU(2)_L$ & 2 & 3 & 3 & 1 & 2 & 2& 2 & 2 \\
\hline
$S_4$ & ${\bf 3}$ & ${\bf 3}$ & ${\bf 1}$ & ${\bf 3}$  & ${\bf 1}$ & ${\bf 2}$ & ${\bf 3}$ & ${\bf 3'}$
 \\
\hline
\end{tabular}
\label{matter-content-S4}
\end{table}
The Lorentz-, gauge- and $S_4$-invariant terms relevant for the neutrino mass matrix are
\bea
\label{Lagrangian-neutrino-S4-antiequality}
{\cal L} &\ni& Y  \left( D^T_{L1} C^{-1} i \tau_2  D_{L1} + D^T_{L2} C^{-1} i \tau_2  D_{L3} +  D^T_{L3} C^{-1} i \tau_2  D_{L2}  \right)  \Delta_4 \nn \\ &&
+ Y' \left[ \left( D^T_{L3} C^{-1} i \tau_2  D_{L3} - D^T_{L2} C^{-1} i \tau_2  D_{L2} \right) \Delta_1 \right.\nn \\ &&
+ \left( D^T_{L1} C^{-1} i \tau_2 D_{L 3} + D^T_{L3} C^{-1} i \tau_2  D_{L1} \right) \Delta_3 \nn \\ &&
\left. - \left( D^T_{L1} C^{-1} i \tau_2 D_{L2} +D^T_{L2} C^{-1} i \tau_2  D_{L1} \right) \Delta_2 \right].
\eea
The $Y(Y')$-term picks up the singlet (triplet) combination from the product of the two triplets ($D^T_{L_i}$ and $D_{L_i}$) (Eq. \ref{3x3-s4}), before multiplying it with the Higgs flavor singlet $\Delta_4$ (triplet $\Delta_i$). We get, upon acquiring small vevs for $\Delta_i^o, i=1,\ldots,4$, the characteristic constraints ($M_{\n33}=-M_{\n22}=Y' \langle \Delta_1^0\rangle$) and ($M_{\n11}=M_{\n23}= Y \langle \Delta_4^0\rangle$).

\subsubsection{Charged lepton sector:}
\label{subsubsection-s4-charged}
We did not find a way to construct a non-degenerate diagonal charged lepton mass matrix $M_{\ell}$. However, we can build a generic mass matrix and impose suitable hierarchy conditions in order to diagonlize $M_\ell$ by rotating infinitesimally the left-handed charged lepton fields. This means that, up to approximations of the order of the charged lepton mass-ratios hierarchies, we are in the `flavor' basis, and the previous phenomenological study is valid. These corrections due to rotating the fields are not larger than other, hitherto discarded, corrections coming, say, from radiative renormalization group running from the seesaw high scale to the observed data low scale.

Noting that $D_{Li}$ transforming under (${\cal D}$) implies that $\overline{D}_{Li}$ would transform under ${\cal D^*}$, one can use Eq. (\ref{3*x3-s4}) of the product (${\bf 3^*} \otimes {\bf 3}$) and get output irreps of (${\bf 1}$), to be multiplied by a Higgs flavor singlet $\phi_I$, and of (${\bf 2}$), to be multiplied by a Higgs flavor doublet $\phi_{II}$ (c.f. Eq. \ref{2x2-s4}), and of (${\bf 3^*}$), to be multiplied by a Higgs flavor triplet $\phi_{III}$ (c.f. Eq. \ref{3*x3-s4}), and finally of (${\bf 3'^*}$), to be multiplied by another Higgs flavor triplet $\phi_{III'}$ (c.f. Eq. \ref{3*x3-s4}).
The relevant Lagrangian is:
\bea
\label{Lagrangian-chargedlepton-S4-anti}
{\cal L} &\ni& \lambda_1  \left( \overline{D}_{L1} \ell_{R1} + \overline{D}_{L2} \ell_{R2} + \overline{D}_{L3} \ell_{R3}  \right) \phi_I  \\  && +
 \lambda_2  \left[ \overline{D}_{L1} \ell_{R1} \phi_{II_1} -\frac{1}{2} \left( \overline{D}_{L2} \ell_{R2} + \overline{D}_{L3} \ell_{R3}  \right)  \phi_{II_1}   +\frac{\sqrt{3}}{2}  \left( \overline{D}_{L2} \ell_{R3} +  \overline{D}_{L3} \ell_{R2} \right) \phi_{II_2} \right] \nn \\ &&+
\lambda_3  \left[ \left( \overline{D}_{L3} \ell_{R2} - \overline{D}_{L2} \ell_{R3} \right) \phi_{III_1} + \left( \overline{D}_{L1} \ell_{R2} + \overline{D}_{L3} \ell_{R1} \right) \phi_{III_2} +\left( -\overline{D}_{L1} \ell_{R3} - \overline{D}_{L2} \ell_{R1} \right) \phi_{III_3} \right] \nn \\  && +
\lambda'_3  \left[ \left( \overline{D}_{L3} \ell_{R3} - \overline{D}_{L2} \ell_{R2} \right) \phi_{III'_1} + \left( -\overline{D}_{L1} \ell_{R3} + \overline{D}_{L2} \ell_{R1} \right) \phi_{III'_2} +\left( \overline{D}_{L1} \ell_{R2} - \overline{D}_{L3} \ell_{R1} \right) \phi_{III'_3} \right], \nn
\eea
which leads, when $\phi_i, i\in\{I,II,III,III'\}$ acquires a vev, to a  charged lepton mass:
\bea
\label{charged-Lepton-mass-matrix-S4-anti}
M_{\ell} &=& \lambda_1 \left( \begin{array}{ccc}\langle \phi_{I}\rangle_0&0&0\\0&\langle \phi_{I}\rangle_0&0\\0&0&\langle \phi_{I}\rangle_0\end{array}\right)
+\lambda_2 \left( \begin{array}{ccc}\langle \phi_{II_1}\rangle_0&0&0\\0&-\frac{1}{2}\langle \phi_{II_1}\rangle_0&\frac{\sqrt{3}}{2}\langle \phi_{II_2}\rangle_0\\0&\frac{\sqrt{3}}{2}\langle \phi_{II_2}\rangle_0&-\frac{1}{2}\langle \phi_{II_1}\rangle_0\end{array}\right)  \\ &&+
\lambda_3 \left( \begin{array}{ccc}0&\langle \phi_{III_2}\rangle_0&-\langle \phi_{III_3}\rangle_0\\-\langle \phi_{III_3}\rangle_0&0&-\langle \phi_{III_1}\rangle_0\\\langle \phi_{III_2}\rangle_0&\langle \phi_{III_1}\rangle_0&0\end{array}\right) +
\lambda'_3 \left( \begin{array}{ccc}0&\langle \phi_{III_3'}\rangle_0&-\langle \phi_{III_2'}\rangle_0\\-\langle \phi_{III_2'}\rangle_0&-\langle \phi_{III_1'}\rangle_0&0\\-\langle \phi_{III_3'}\rangle_0&0&\langle \phi_{III_1'}\rangle_0\end{array}\right) \nn
\eea
Two common ways to get a generic $M_\ell$.
\begin{itemize}
\item We assume a vev hierarchy such that the $S_4$-first components are dominant and comparable ($\langle \phi_I\rangle_0 \approx \langle \phi_{II_1}\rangle_0 \approx \langle \phi_{III_1}\rangle_0 \approx \langle \phi_{III_1'}\rangle_0 \approx v$, whereas other VEVs can be discarded). We do not study the Higgs scalar potential, but assume that its various free parameters can be adjusted so that to lead naturally to this assumption.  This implies a diagonal $M_\ell$:
    \bea
    \label{m-ell-first}
    M_\ell &\approx & v \mbox{ diag} \left(\lambda_1+\lambda_2,\;\;\; \lambda_1-\frac{1}{2}\lambda_2-\lambda'_3, \;\;\; \lambda_1-\frac{1}{2}\lambda_2+\lambda'_3\right)
    \eea
The mass matrix is approximately diagonal with enough parameters to produce the observed charged lepton mass hierarchies by taking:
\bea
&m_e \approx (\lambda_1+\lambda_2) v, \;\;\;m_\mu \approx (\lambda_1-\frac{1}{2}\lambda_2-\lambda'_3)v , \;\;\; m_\tau \approx (\lambda_1-\frac{1}{2}\lambda_2+\lambda'_3)v.
\eea
So, we are, up to a good approximation which can be adjusted to be of the order of the mass ratio $\leq 10^{-2}$, in the flavor basis. The effect of the ``small" neglected non-diagonal terms is to require rotating infinitesimally the left handed charged lepton fields, leading thus to corrections on the observed $V_{\mbox{\tiny PMNS}}$ of the same small order $10^{-2}$.
\item Looking at Eq. (\ref{charged-Lepton-mass-matrix-S4-anti}), we see that we have 9 free vevs and 4 free perturbative coupling constants, appearing in 9 linear combinations, {\it a priori} enough to construct the generic $3 \times 3$ complex matrix. Thus,  $M_\ell$ can be casted in the form
   \bea
\label{M_elltype2}
M_{\ell } =  \left( \begin {array}{c}
{\bf a}^T\\{\bf b}^T\\{\bf c}^T
\end {array}
\right) &\Rightarrow&
M_{\ell } M_{\ell}^\dagger = \left(\begin {array}{ccc}
{\bf a.a} &{\bf a.b}&{\bf a.c} \\
{\bf b.a} &{\bf b.b}&{\bf b.c}\\
{\bf c.a} &{\bf c.b}&{\bf c.c}
\end {array} \right)
\eea
where ${\bf a}, {\bf b}$ and ${\bf c}$ are three linearly independent vectors, so taking only the following natural assumption on the norms of the vectors
\bea \parallel {\bf a} \parallel /\parallel {\bf c} \parallel = m_e/m_\tau \sim 3 \times 10^{-4} &,&  \parallel {\bf b} \parallel /\parallel {\bf c} \parallel = m_\mu/m_\tau \sim 6 \times 10^{-2}\eea
one can diagonalize $M_{\ell } M_{\ell}^\dagger$ by an infinitesimal rotation as was done in \cite{0texture}, which proves that we are to a good approximation in the flavor basis.

\end{itemize}

\section{Summary and Conclusion}
In this study, we carry out a systematic study of three textures defined by two constraints (one equality and one anti-equality). These textures are realizable through the non-abelian group $S_4$. We compute the analytical expressions of the A's and B's (Eq. \ref{Coff}) in the general case where all $12$ parameters of the mass matrix are considered, including the unphysical phases.

We delved into the subtle intricacies the unphysical phases have within the texture studies, although they can be absorbed by the charged lepton fields. In particular, we single out the role of the unphysical phases in the definition of any texture, so that all past studies restricted to vanishing unphysical phases case should be looked upon as studies of textures defined not merely by a mathematical constraint, but rather via a mathematical constraint defined on the slice of vanishing unphysical phases, which, in their turn, depend on the parametrization adopted in the analysis.

We discussed three different definitions of a given texture, that we called ``Mathematical'', ``Specific" and ``Generalized''. The `non-physicality' of the first (second) definition originates from it being sensitive to the unphysical phases (PMNS parametrization). Only the third definition is insensitive to the unphysical phases and also is independent of the PMNS parametrization. Moreover, in model building, the first and the third definitions are relatively easier to realize than the second. Regarding phenomenology, and putting aside the unphysical phases, the first and third definitions are equivalent.

We carried out a complete phenomenological analysis, adopting first the vanishing unphysical phases scenario. Here, the analytical formulae for the mass ratios are relatively simple, and we stated the leading order terms in $s_z$ for the neutrino physical parameter $R_\n$ and other measurable parameters. Moreover, we could interpret analytically the resulting correlations, by following a simplifying strategy of equating $R_\n$ to zero and looking, numerically and analytically, at the resulting correlations which are similar to the real correlations. We found that all three textures accommodate the {\bf IH} ordering, but the texture III also accommodates, in a very narrow band around $\t_{23}\equiv\t_y \approx 50^o$, the {\bf NH} ordering as well.

When including the unphysical phases, we find that many correlations get ``diluted" (voire ``disappeared''), since, with these unphysical phases not necessarily zero, one can find many more points in the parameter space meeting the experimental constraints and the mathematical condition defining the texture. We present the relevant correlations, and note that now all three textures can accommodate the two types of hierarchy {\bf IH \& NH}. We stress that the differing phenomenologies do not imply any physical effects for the unphysical phases. Rather, the change in phenomenology actually stems from the different textures that we need to define in a consistent way insensitive to the unphysical phases.

Adopting the stringent cosmological bound of (Eq. \ref{cosTB}) shows clearly the ``phenomenological" role played by the unphysical phases, in that no texture under study was viable when these phases were put to zero, whereas switching on the unphysical phases allowed one texture at {\bf NH} ordering to survive the experimental tight bounds.

Finally, we presented a theoretical realization of these textures via a type-II seesaw scenario and assuming a non-abelian group $S_4$.    However, we have not discussed the question of the scalar potential
and finding its general form under the imposed symmetry. Nor did we deal with the radiative corrections effect on the phenomenology and whether or not it can spoil the form of the texture while running from the seesaw “ultraviolet” scale, where the mass matrix is bound to take its texture form, to the low scale where phenomenology was analyzed.

\section*{{\large \bf Acknowledgements}}
E. I. L acknowledges support from ICTP through the Senior Associate program. N. C. acknowledges support from the CAS PIFI fellowship and from the Humboldt Foundation. E. L.'s work was partially supported by the STDF project 37272.

\vspace{1cm}
\appendix

\section{{\large \bf Appendix: $S_4$-irreps}}
\label{appendix-s4}

We denote by $n_r$ ($m_n$) the number of (of $n$-dim) inequivalent unitary irreps for a finite group of cardinality $n_G$ with $n_c$ conjugation equivalence classes. With $\chi_\a$ denoting the character of the irrep $\a$, we have the following rules:
\bea
  n_c = n_r &,& \sum_{n\in N} m_n n^2 = n_G, \label{irreps-count}\\
  \sum_{a\in G} \chi_\a(a) \chi_\b^*(a) &=& n_G \d_{\a\b}, \label{orthogonality} \\
  \sum_{\a \in \mbox{irreps}} \chi_\a(a) \chi_\a^*(b) &=& \frac{n_G}{\mbox{card}[a]} \d_{[a][b]} \label{orthogonality_tr}.
\eea

The symmetry group of order $4$, $S_4$, is the group of permutations of $\{1,2,3,4\}$. It has 24 elements, given in terms of cycles' notation as: $S_4 = \{a_1=e,a_2=(12)(34), a_3=(13)(24), a_4=(14)(23), b_1=(243), b_2=(142), b_3=(123), b_4=(134), c_1=(234), c_2=(132), c_3=(143), c_4=(124), d_1 =(34), d_2=(12), d_3=(1423), d_4=(1324), e_1=(23), e_2=(1342), e_3=(1243), e_4=(14), f_1=(24), f_2=(1432), f_3=(13), f_4=(1234)\}$, which can be divided into five classes (the class $s C_h=[g]$ includes $s$ elements $g$ all of order $h$\footnote{the order of an element $g$ is the order (cardinality) of the subgroup generated by this element and is equal to ($\min\{n\in N\backslash\{0\}: g^n=1\}$). For a permutation written as a product of disjoint cycles, the order is the least common multiplier of the cardinalities of these cycles' supports.})
\bea & 1C_1=\{e\}, \; \; 3C_2=\{a_2, a_3, a_4\}, \;\;\; 6C_2=\{d_1, d_2, e_1, e_4, f_1, f_3\},  \nn \\  &  8C_3=\{b_1, b_2, b_3, b_4, c_1, c_2, c_3, c_4\}, \;\;\;
 6C_4=\{d_3, d_4, e_2, e_3, f_2, f_4\},\eea

 $S_4$ has two generators, and is equivalently defined as :
\bea
S_4 &=& \langle D, B: D^4=B^3=1,DB^2D=B\rangle ,\nn \\
&=& \langle T, S: T^4=S^2=(ST)^3=1 \rangle,
\eea
with the first (second) definition leading to $DBD=BD^2B$ ($(TS)^3=1$). One can take ($D=d_4, B= b_1$) for the first set of generators, or ($T=D, S=BD^{-1}$) for the second set.

Thus, by applying Eqs. (\ref{irreps-count}), we have five unitary inequivalent irreps, and:
\bea \sum_{n\in N}m_n=5, \sum_{n\in N}m_nn^2=24 &\Rightarrow&  m_1=2, m_2=1, m_3=2\eea
Applying Eqs. (\ref{orthogonality}, \ref{orthogonality_tr}), we have the character table of $S_4$ (Table \ref{characterTableS4}).
\begin{table}[h]
\caption{Character table of $S_4$}
\centering
\begin{tabular}{cccccc}
\hline
\hline
classes$/$irreps & $\chi_{\bf 1}$ & $\chi_{\bf 1'}$ & $\chi_{\bf 2}$ & $\chi_{\bf 3}$ & $\chi_{\bf 3'}$  \\
\hline
\hline
$1C_1$ & 1 &1 & 2 & 3 & 3 \\
\hline
$3C_2$ & 1 & 1  & 2 & -1 & -1 \\
\hline
$6C_2$ & 1 & -1  & 0 & 1 & -1 \\
\hline
$6C_4$ & 1 & -1  & 0 & -1 & 1 \\
\hline
$8C_3$ & 1 & 1  & -1 & 0 & 0 \\
\hline
\end{tabular}
\label{characterTableS4}
\end{table}

In the canonical basis ($a_1, a_2, a_3, a_4$), the linear combination ($a_1+a_2+a_3+a_4$) is invariant under the action of the permutations representation. Thus, the orthogonal subsapce spanned by
\bea
\label{A-space}
\left( \begin{array}{c} A_x \\ A_y\\ A_z\end{array} \right) &=& \left( \begin{array}{c} a_1 +a_2-a_3-a_4 \\ a_1 -a_2 + a_3 -a_4 \\ a_1 -a_2 -a_3 + a_4\end{array}\right)
\eea
is also invariant. The restriction of the permutations representation onto the $3$-dim $A$-space is the ${\bf 3}$ irrep given, in this $A$-basis, by:
\bea
\label{A-basis-3}
b_1^{\mbox{\tiny can.}} = \left(\begin{array}{cccc} 1&0&0&0\\ 0&0&0&1\\0&1&0&0 \\0&0&1&0 \end{array}\right) &\Rightarrow& b_1^{\mbox{\tiny A}} = \left(\begin{array}{ccc} 0&0&1\\ 1&0&0\\0&1&0 \end{array}\right) \\
d_4^{\mbox{\tiny can.}} = \left(\begin{array}{cccc} 0&0&1&0\\ 0&0&0&1\\0&1&0&0\\ 1&0&0&0 \end{array}\right) &\Rightarrow& d_4^{\mbox{\tiny A}} = \left(\begin{array}{ccc} -1&0&0\\ 0&0&-1\\0&1&0 \end{array}\right),
\eea
whereas the ${\bf 3'}$ irrep is given, in an $A$-like basis, by:
\bea
 {b'}_1^{\mbox{\tiny A}} = \left(\begin{array}{ccc} 0&0&1\\ 1&0&0\\0&1&0 \end{array}\right) &,&
{d'}_4^{\mbox{\tiny A}} = \left(\begin{array}{ccc} 1&0&0\\ 0&0&1\\0&-1&0 \end{array}\right) \label{A-basis-3'},
\eea
and the ${\bf 2}$ irrep is:
\bea
 {b''}_1^{\mbox{\tiny A}} = \left(\begin{array}{cc} \omega&0\\ 0&\omega^2 \end{array}\right) &,&
{d''}_4^{\mbox{\tiny A}} = \left(\begin{array}{cc} 0&1\\ 1&0 \end{array}\right) \label{A-basis-2},
\eea
where $\omega = e^{i2\pi/3}$, and one can compute the corresponding $T^{\mbox{\tiny(','')A}} =d_4^{\mbox{\tiny(','')A}}, S^{\mbox{\tiny(','')A}}: S^{\mbox{\tiny(','')A}}T^{\mbox{ \tiny(','')A}}=b_1^{\mbox{\tiny(','')A}}$ in these irreps.

Another common basis is the ${B}$-basis given by the unitary similarity matrices $U_{\mbox{\tiny doublet}}, U_{\mbox{\tiny triplet}}$:
\bea
\label{s4-from-A-to-tilde}
U_{\mbox{\tiny doublet}}= \frac{1}{\sqrt{2}} \left( \begin{array}{cc}1&i\\1&-i\end{array}\right) &,&
U_{\mbox{\tiny triplet}}= \frac{1}{\sqrt{2}} \left( \begin{array}{ccc}\sqrt{2}&0&0\\0&1&1 \\ 0& i& -i\end{array}\right),
\eea
so we have:
\bea
b_1   =U_{\mbox{\tiny triplet}}^\dagger b_1^{\mbox{\tiny  A}} U_{\mbox{\tiny triplet}} =\left( \begin{array}{ccc}0 &\frac{i}{\sqrt{2}}&\frac{-i}{\sqrt{2}} \\ \frac{1}{\sqrt{2}}&\frac{-i}{2}&\frac{-i}{2}\\ \frac{1}{\sqrt{2}}&\frac{i}{2}& \frac{i}{2}\end{array}\right) &,&
d_4   =U_{\mbox{\tiny triplet}}^\dagger d_4^{\mbox{\tiny  A}} U_{\mbox{\tiny triplet}} =\mbox{diag}\left(-1,-i,i\right) ,
\nn \\
b'_1   =U_{\mbox{\tiny triplet}}^\dagger {b'}_1^{\mbox{\tiny  A}} U_{\mbox{\tiny triplet}} =\left( \begin{array}{ccc} 0&\frac{i}{\sqrt{2}}&\frac{-i}{\sqrt{2}} \\ \frac{1}{\sqrt{2}}&\frac{-i}{2}&\frac{-i}{2} \\ \frac{1}{\sqrt{2}}&\frac{i}{2}&\frac{i}{2} \end{array}\right) &,&
d'_4   =U_{\mbox{\tiny triplet}}^\dagger {d'}_4^{\mbox{\tiny  A}} U_{\mbox{\tiny triplet}} =\mbox{diag}\left(1,i,-i\right) ,\nn \\
b''_1   =U_{\mbox{\tiny doublet}}^\dagger {b''}_1^{\mbox{\tiny  A}} U_{\mbox{\tiny doublet}} =\frac{1}{2}\left( \begin{array}{ccc} -1&-\sqrt{3} \\ \sqrt{3}&-1 \end{array}\right) &,&
d''_4   =U_{\mbox{\tiny doublet}}^\dagger {d''}_4^{\mbox{\tiny  A}} U_{\mbox{\tiny doublet}} =\mbox{diag}\left(1,-1\right) ,\label{s4-tilde-basis-2}
\eea
and one can compute $T,T',T'',S,S', S''$.

Referring to \cite{Ishimori_2010} for details, we state explicitly in appendix (\ref{appendix_mult}) the irreps multiplication rules in the $B$-basis adopted to define the texture and the matter fields symmetry assignments.

\section{{$S_4$ multiplication rules in the working-base:}}
\label{appendix_mult}
The symmetric group of order $4$ has two generators, and can be defined minimally as:
\bea
S_4 &=& \langle d, b: d^4=b^3=1,db^2d=b\rangle
= \langle T, S: T^4=S^2=(ST)^3=1 \rangle ,
\eea
leading to ($dbd=bd^2b$, and to $(TS)^3=1$), and where one can take ($T=d, ST=b$) linking the two sets of two-generators. $S_4$ has five inequivalent irreps (${\bf 1}, {\bf 1'}, {\bf 2}, {\bf 3}$ and ${\bf 3'}$). In appendix (\ref{appendix-s4}), we stated the expressions of the generators in a certain working ${B}$-basis, where the symmetry assignments for the matter fields are given, and where the texture of the mass matrix is of the required form. Thus we have ($d^{(','')},b^{(','')}$ refer to ${\bf 3}({\bf 3'}, {\bf 2})$) (c.f. Eqs. \ref{s4-tilde-basis-2}):
\bea
\label{working-basis-s4}
d=\mbox{diag}(-1,-i,i) &,& b= \left( \begin{array}{ccc}0 &\frac{i}{\sqrt{2}}&\frac{-i}{\sqrt{2}} \\ \frac{1}{\sqrt{2}}&\frac{-i}{2}&\frac{-i}{2}\\ \frac{1}{\sqrt{2}}&\frac{i}{2}& \frac{i}{2}\end{array}\right), \\
d'=\mbox{diag}(1,i,-i) &,& b'= \left( \begin{array}{ccc}0 &\frac{i}{\sqrt{2}}&\frac{-i}{\sqrt{2}} \\ \frac{1}{\sqrt{2}}&\frac{-i}{2}&\frac{-i}{2}\\ \frac{1}{\sqrt{2}}&\frac{i}{2}& \frac{i}{2}\end{array}\right), \\
d''=\mbox{diag}(1,-1) &,& b''=\frac{1}{2}\left( \begin{array}{ccc} -1&-\sqrt{3} \\ \sqrt{3}&-1 \end{array}\right).
\eea
One can then check that the following ``symmetry adapted linear combinations" (S.A.L.C.) multiplication rules are valid in the adopted working $B$-basis.
\bea
\label{2x2-s4}
\left( \begin{array}{c} x_1 \\ x_2 \end{array}\right)_{\bf 2} \otimes \left( \begin{array}{c} y_1 \\ y_2\end{array}\right)_{\bf 2} &=& \left(x_1y_1+x_2y_2\right)_{\bf 1} \oplus \left( x_1y_2-x_2y_1 \right)_{\bf 1'} \oplus \left( \begin{array}{c}x_2y_2-x_1y_1 \\ x_1y_2+x_2y_1 \end{array} \right)_{\bf 2},
\eea

\bea
\label{2x3-s4}
\left(\begin{array}{l}
x_{1} \\
x_{2}
\end{array}\right)_{\bf 2} \otimes\left(\begin{array}{l}
y_{1} \\
y_{2} \\
y_{3}
\end{array}\right)_{\bf 3}&=&\left(\begin{array}{c}
x_{1} y_{1} \\
\frac{\sqrt{3}}{2} x_{2} y_{3}-\frac{1}{2} x_{1} y_{2} \\
\frac{\sqrt{3}}{2} x_{2} y_{2}-\frac{1}{2} x_{1} y_{3}
\end{array}\right)_{\bf 3} \oplus\left(\begin{array}{c}
-x_{2} y_{1} \\
\frac{\sqrt{3}}{2} x_{1} y_{3}+\frac{1}{2} x_{2} y_{2} \\
\frac{\sqrt{3}}{2} x_{1} y_{2}+\frac{1}{2} x_{2} y_{3}
\end{array}\right)_{\bf 3'},\\
\label{2x3'-s4}
\left(\begin{array}{l}
x_{1} \\
x_{2}
\end{array}\right)_{\bf 2} \otimes\left(\begin{array}{l}
y_{1} \\
y_{2} \\
y_{3}
\end{array}\right)_{\bf 3'}&=&\left(\begin{array}{c}
-x_{2} y_{1} \\
\frac{\sqrt{3}}{2} x_{1} y_{3}+\frac{1}{2} x_{2} y_{2} \\
\frac{\sqrt{3}}{2} x_{1} y_{2}+\frac{1}{2} x_{2} y_{3}
\end{array}\right)_{\bf 3} \oplus\left(\begin{array}{c}
x_{1} y_{1} \\
\frac{\sqrt{3}}{2} x_{2} y_{3}-\frac{1}{2} x_{1} y_{2} \\
\frac{\sqrt{3}}{2} x_{2} y_{2}-\frac{1}{2} x_{1} y_{3}
\end{array}\right)_{\bf 3'} \text {, }
\eea

\bea
\left(\begin{array}{l}
x_{1} \\
x_{2} \\
x_{3}
\end{array}\right)_{\bf 3(3')} \otimes\left(\begin{array}{l}
y_{1} \\
y_{2} \\
y_{3}
\end{array}\right)_{\bf 3 (3')}&=& \left(x_{1} y_{1}+x_{2} y_{3}+x_{3} y_{2}\right)_{\bf 1} \oplus \left(\begin{array}{c}
x_{1} y_{1}-\frac{1}{2}\left(x_{2} y_{3}+x_{3} y_{2}\right) \\
\frac{\sqrt{3}}{2}\left(x_{2} y_{2}+x_{3} y_{3}\right)
\end{array}\right)_{\bf 2} \nn \\
&& \oplus\left(\begin{array}{c}
x_{3} y_{3}-x_{2} y_{2} \\
x_{1} y_{3}+x_{3} y_{1} \\
-x_{1} y_{2}-x_{2} y_{1}
\end{array}\right)_{\bf 3} \oplus\left(\begin{array}{c}
x_{3} y_{2}-x_{2} y_{3} \\
-x_{1} y_{2}+x_{2} y_{1} \\
x_{1} y_{3}-x_{3} y_{1}
\end{array}\right)_{\bf 3'}, \label{3x3-s4} \\
\left(\begin{array}{l}
x_{1} \\
x_{2} \\
x_{3}
\end{array}\right)_{\bf 3} \otimes\left(\begin{array}{l}
y_{1} \\
y_{2} \\
y_{3}
\end{array}\right)_{\bf 3'}&=&\left(x_{1} y_{1}+x_{2} y_{3}+x_{3} y_{2}\right)_{\bf 1'}
\oplus\left(\begin{array}{c} \frac{\sqrt{3}}{2}\left(x_{2} y_{2}+x_{3} y_{3}\right) \\ -x_{1} y_{1}+\frac{1}{2}\left(x_{2} y_{3}+x_{3} y_{2}\right)\end{array}\right)_{\bf 2} \nn\\
&&\oplus\left(\begin{array}{l}
x_{3} y_{2}-x_{2} y_{3} \\
-x_{1} y_{2}+x_{2} y_{1} \\
x_{1} y_{3}-x_{3} y_{1}
\end{array}\right)_{\bf 3} \oplus\left(\begin{array}{c}
x_{3} y_{3}-x_{2} y_{2} \\
+x_{1} y_{3}+x_{3} y_{1} \\
-x_{1} y_{2}-x_{2} y_{1}
\end{array}\right)_{\bf 3'},  \label{3x3'-s4}
\eea
We state also the rules involving conjugate irreps ($\cal D^*$ which is equivalent to $\cal D$):
\bea
\label{2*x2-s4}
\left( \begin{array}{c} x^*_1 \\ x^*_2 \end{array}\right)_{\bf 2^*} \otimes \left( \begin{array}{c} y_1 \\ y_2\end{array}\right)_{\bf 2} &=& \left(x^*_1y_1+x^*_2y_2\right)_{\bf 1} \oplus \left( x^*_1y_2-x^*_2y_1 \right)_{\bf 1'} \oplus \left( \begin{array}{c}x^*_2y_2-x^*_1y_1 \\ x^*_1y_2+x^*_2y_1 \end{array} \right)_{\bf 2},
\eea
\bea
\label{2*x3-s4}
\left(\begin{array}{l}
x^*_{1} \\
x^*_{2}
\end{array}\right)_{\bf 2^*} \otimes\left(\begin{array}{l}
y_{1} \\
y_{2} \\
y_{3}
\end{array}\right)_{\bf 3}&=&\left(\begin{array}{c}
x^*_{1} y_{1} \\
\frac{\sqrt{3}}{2} x^*_{2} y_{3}-\frac{1}{2} x^*_{1} y_{2} \\
\frac{\sqrt{3}}{2} x^*_{2} y_{2}-\frac{1}{2}x^*_{1} y_{3}
\end{array}\right)_{\bf 3} \oplus\left(\begin{array}{c}
-x^*_{2} y_{1} \\
\frac{\sqrt{3}}{2} x^*_{1} y_{3}+\frac{1}{2} x^*_{2} y_{2} \\
\frac{\sqrt{3}}{2} x^*_{1} y_{2}+\frac{1}{2} x^*_{2} y_{3}
\end{array}\right)_{\bf 3'},\\
\label{2*x3'-s4}
\left(\begin{array}{l}
x^*_{1} \\
x^*_{2}
\end{array}\right)_{\bf 2^*} \otimes\left(\begin{array}{l}
y_{1} \\
y_{2} \\
y_{3}
\end{array}\right)_{\bf 3'}&=&\left(\begin{array}{c}
-x^*_{2} y_{1} \\
\frac{\sqrt{3}}{2} x^*_{1} y_{3}+\frac{1}{2} x^*_{2} y_{2} \\
\frac{\sqrt{3}}{2} x^*_{1} y_{2}+\frac{1}{2} x^*_{2} y_{3}
\end{array}\right)_{\bf 3} \oplus\left(\begin{array}{c}
x^*_{1} y_{1} \\
\frac{\sqrt{3}}{2} x^*_{2} y_{3}-\frac{1}{2} x^*_{1} y_{2} \\
\frac{\sqrt{3}}{2} x^*_{2} y_{2}-\frac{1}{2}x^*_{1} y_{3}
\end{array}\right)_{\bf 3'} \text {, }
\eea

\bea
\label{3*x3-s4}
\left(\begin{array}{l}
x^*_{1} \\
x^*_{2} \\
x^*_{3}
\end{array}\right)_{\bf 3^*(3'^*)} \otimes\left(\begin{array}{l}
y_{1} \\
y_{2} \\
y_{3}
\end{array}\right)_{\bf 3 (3')}&=& \left(x^*_{1} y_{1}+x^*_{2} y_{2}+x^*_{3} y_{3}\right)_{\bf 1} \oplus \left(\begin{array}{c}
x^*_{1} y_{1}-\frac{1}{2}\left(x^*_{2} y_{2}+x^*_{3} y_{3}\right) \\
\frac{\sqrt{3}}{2}\left(x^*_{2} y_{3}+x^*_{3} y_{2}\right)
\end{array}\right)_{\bf 2} \nn \\
&& \oplus\left(\begin{array}{c}
x^*_{3} y_{2}-x^*_{2} y_{3} \\
x^*_{1} y_{2}+x^*_{3} y_{1} \\
-x^*_{1} y_{3}-x^*_{2} y_{1}
\end{array}\right)_{\bf 3^*} \oplus\left(\begin{array}{c}
x^*_{3} y_{3}-x^*_{2} y_{2} \\
-x^*_{1} y_{3}+x^*_{2} y_{1} \\
x^*_{1} y_{2}-x^*_{3} y_{1}
\end{array}\right)_{\bf 3'^*},
\eea
\bea
\label{3*x3'-s4}
\left(\begin{array}{l}
x^*_{1} \\
x^*_{2} \\
x^*_{3}
\end{array}\right)_{\bf 3^*} \otimes\left(\begin{array}{l}
y_{1} \\
y_{2} \\
y_{3}
\end{array}\right)_{\bf 3'}&=& \left(x^*_{1} y_{1}+x^*_{2} y_{2}+x^*_{3} y_{3}\right)_{\bf 1'} \oplus \left(\begin{array}{c}
\frac{\sqrt{3}}{2}\left(x^*_{2} y_{3}+x^*_{3} y_{2}\right) \\-x^*_{1} y_{1}+\frac{1}{2}\left(x^*_{2} y_{2}+x^*_{3} y_{3}\right)
\end{array}\right)_{\bf 2} \nn \\
&& \oplus\left(\begin{array}{c}
x^*_{3} y_{3}-x^*_{2} y_{2} \\
x^*_{2} y_{1}-x^*_{1} y_{3} \\
x^*_{1} y_{2}-x^*_{3} y_{1}
\end{array}\right)_{\bf 3^*} \oplus\left(\begin{array}{c}
x^*_{3} y_{2}-x^*_{2} y_{3} \\
x^*_{1} y_{2}+x^*_{3} y_{1} \\
-x^*_{1} y_{3}-x^*_{2} y_{1}
\end{array}\right)_{\bf 3'^*},
\eea

\bea
\label{s4-2*x2*}
\left( \begin{array}{c} x^*_1 \\ x^*_2 \end{array}\right)_{\bf 2^*} \otimes \left( \begin{array}{c} y^*_1 \\ y^*_2\end{array}\right)_{\bf 2^*} &=& \left(x^*_1y^*_1+x^*_2y^*_2\right)_{\bf 1^*} \oplus \left( x^*_1y^*_2-x^*_2y^*_1 \right)_{\bf 1'^*} \oplus \left( \begin{array}{c}x^*_2y^*_2-x^*_1y^*_1 \\ x^*_1y^*_2+x^*_2y^*_1 \end{array} \right)_{\bf 2^*}
\eea

\bea
\label{2*x3*-s4}
\left(\begin{array}{l}
x^*_{1} \\
x^*_{2}
\end{array}\right)_{\bf 2^*} \otimes\left(\begin{array}{l}
y^*_{1} \\
y^*_{2} \\
y^*_{3}
\end{array}\right)_{\bf 3^*}&=&\left(\begin{array}{c}
x^*_{1} y^*_{1} \\
\frac{\sqrt{3}}{2} x^*_{2} y^*_{3}-\frac{1}{2} x^*_{1} y^*_{2} \\
\frac{\sqrt{3}}{2} x^*_{2} y^*_{2}-\frac{1}{2}x^*_{1} y^*_{3}
\end{array}\right)_{\bf 3^*} \oplus\left(\begin{array}{c}
-x^*_{2} y^*_{1} \\
\frac{\sqrt{3}}{2} x^*_{1} y^*_{3}+\frac{1}{2} x^*_{2} y^*_{2} \\
\frac{\sqrt{3}}{2} x^*_{1} y^*_{2}+\frac{1}{2} x^*_{2} y^*_{3}
\end{array}\right)_{\bf 3'^*},\\
\label{2*x3'*-s4}
\left(\begin{array}{l}
x^*_{1} \\
x^*_{2}
\end{array}\right)_{\bf 2^*} \otimes\left(\begin{array}{l}
y^*_{1} \\
y^*_{2} \\
y^*_{3}
\end{array}\right)_{\bf 3'^*}&=&\left(\begin{array}{c}
-x^*_{2} y^*_{1} \\
\frac{\sqrt{3}}{2} x^*_{1} y^*_{3}+\frac{1}{2} x^*_{2} y^*_{2} \\
\frac{\sqrt{3}}{2} x^*_{1} y^*_{2}+\frac{1}{2} x^*_{2} y^*_{3}
\end{array}\right)_{\bf 3^*} \oplus\left(\begin{array}{c}
x^*_{1} y^*_{1} \\
\frac{\sqrt{3}}{2} x^*_{2} y^*_{3}-\frac{1}{2} x^*_{1} y^*_{2} \\
\frac{\sqrt{3}}{2} x^*_{2} y^*_{2}-\frac{1}{2}x^*_{1} y^*_{3}
\end{array}\right)_{\bf 3'^*} \text {, }
\eea

\bea
\left(\begin{array}{l}
x^*_{1} \\
x^*_{2} \\
x^*_{3}
\end{array}\right)_{\bf 3^*(3'^*)} \otimes\left(\begin{array}{l}
y^*_{1} \\
y^*_{2} \\
y^*_{3}
\end{array}\right)_{\bf 3^* (3'^*)}&=& \left(x^*_{1} y^*_{1}+x^*_{2} y^*_{3}+x^*_{3} y^*_{2}\right)_{\bf 1^*} \oplus \left(\begin{array}{c}
x^*_{1} y^*_{1}-\frac{1}{2}\left(x^*_{2} y^*_{3}+x^*_{3} y^*_{2}\right) \\
\frac{\sqrt{3}}{2}\left(x^*_{2} y^*_{2}+x^*_{3} y^*_{3}\right)
\end{array}\right)_{\bf 2^*} \nn \\
&& \oplus\left(\begin{array}{c}
x^*_{3} y^*_{3}-x^*_{2} y^*_{2} \\
x^*_{1} y^*_{3}+x^*_{3} y^*_{1} \\
-x^*_{1} y^*_{2}-x^*_{2} y^*_{1}
\end{array}\right)_{\bf 3^*} \oplus\left(\begin{array}{c}
x^*_{3} y^*_{2}-x^*_{2} y^*_{3} \\
-x^*_{1} y^*_{2}+x^*_{2} y^*_{1} \\
x^*_{1} y^*_{3}-x^*_{3} y^*_{1}
\end{array}\right)_{\bf 3'^*},  \label{3*x3*-s4} \\
\left(\begin{array}{l}
x^*_{1} \\
x^*_{2} \\
x^*_{3}
\end{array}\right)_{\bf 3^*} \otimes\left(\begin{array}{l}
y^*_{1} \\
y^*_{2} \\
y^*_{3}
\end{array}\right)_{\bf 3'^*}&=&\left(x^*_{1} y^*_{1}+x^*_{2} y^*_{3}+x^*_{3} y^*_{2}\right)_{\bf 1'^*}
\oplus\left(\begin{array}{c} \frac{\sqrt{3}}{2}\left(x^*_{2} y^*_{2}+x^*_{3} y^*_{3}\right) \\ -x^*_{1} y^*_{1}+\frac{1}{2}\left(x^*_{2} y^*_{3}+x^*_{3} y^*_{2}\right)\end{array}\right)_{\bf 2^*} \nn\\
&&\oplus\left(\begin{array}{l}
x^*_{3} y^*_{2}-x^*_{2} y^*_{3} \\
x^*_{2} y^*_{1}-x^*_{1} y^*_{2} \\
x^*_{1} y^*_{3}-x^*_{3} y^*_{1}
\end{array}\right)_{\bf 3^*} \oplus\left(\begin{array}{c}
x^*_{3} y^*_{3}-x^*_{2} y^*_{2} \\
x^*_{1} y^*_{3}+x^*_{3} y^*_{1} \\
-x^*_{1} y^*_{2}-x^*_{2} y^*_{1}
\end{array}\right)_{\bf 3'^*},  \label{3*x3'*-s4}
\eea

\bibliographystyle{unsrt}

\end{document}